\documentclass[journal]{IEEEtran}
\usepackage{placeins}
\usepackage{diagbox}
\usepackage[inline]{enumitem}
\usepackage{xcolor}
\usepackage{amsmath,amsfonts}
\usepackage{amsmath,amssymb}
\usepackage{algorithm}
\usepackage[group-separator={,},group-minimum-digits=4]{siunitx}
\usepackage{threeparttable}
\usepackage{array}
\usepackage[caption=false,font=normalsize,labelfont=sf,textfont=sf]{subfig}
\usepackage{textcomp}
\usepackage{microtype}
\usepackage{stfloats}
\usepackage{url}
\usepackage{verbatim}
\usepackage{graphicx}
\usepackage{algpseudocode}
\usepackage{cite}
\usepackage{balance}
\usepackage{mathrsfs}
\usepackage[colorlinks=true,
            linkcolor=blue,
            anchorcolor=blue,
            citecolor=blue]{hyperref}

\begin{document}

\title{
    {Covariance-Weighted Spectral Delay Fusion With a One-Dimensional Affine Model for High-Precision Distributed Optical-Fiber Sensing}
    }
\author{ 
{
    Zhiyang~Xue,
    Huan~Huang, 
    Ziang~Chen,
    Zhongxing~Tian,
    Zeyu~Feng,
    Yuhan~Jiang,
    Dongdong~Zou,
    Jun~Li,
    Gangxiang~Shen,~\textit{Senior~Member,~IEEE,~Fellow,~Optica},
    and Yi~Cai,~\textit{Senior~Member,~IEEE,~Fellow,~Optica}
}
\thanks{  

Z.~Xue, H.~Huang, Z.~Chen, Z.~Tian, Z.~Feng, Y.~Jiang, D.~Zou, J.~Li,  G.~Shen, and Y.~Cai are with the School of Electronic and Information Engineering, Soochow University, Suzhou, Jiangsu 215006, China 
\urlstyle{same}(e-mail: \nolinkurl{zyxue999@stu.suda.edu.cn}; \nolinkurl{hhuang1799@gmail.com}; \nolinkurl{zachen@stu.suda.edu.cn}; \nolinkurl{zxtian@ieee.org}; \nolinkurl{zyfeng1211@stu.suda.edu.cn}; \nolinkurl{yuhanjiangsuda@outlook.com}; \nolinkurl{ddzou@suda.edu.cn}; \nolinkurl{ljun@suda.edu.cn};  \nolinkurl{shengx@suda.edu.cn}; \mbox{\nolinkurl{yicai@suda.edu.cn}}).   
}
}
\maketitle

\begin{abstract}
Periodic disturbances can produce ambiguous delay estimates, limiting reliable high-precision localization in distributed optical-fiber sensing. We develop spectral delay fusion for a sensing system using a dual-wavelength bidirectional Mach--Zehnder interferometer, with four phase traces recovered by heterodyne detection and digital demodulation. With calibrated propagation parameters and timing offsets fixed, the six pairwise delay predictions form a one-dimensional affine line segment parameterized by the position of a single dominant disturbance, with sensitivities determined by propagation direction and chromatic dispersion. A generalized least-squares estimator combines unwrapped delays from robust cross-spectral phase slopes with wrapped delays from polarity-invariant phase alignment to jointly estimate position and integer ambiguities under the proposed model, using an effective joint covariance to account for shared-channel and cross-representation dependence. Experiments use a 131.335-km sensing fiber at 1530 and 1550~nm, with periodic phase perturbations applied at five nominal positions from 25 to 125~km. Across the reported groups of 20 records, the proposed method yields sample standard deviations of 1.007--1.685~m at a drive voltage of 500~mV and 0.449--1.324~m at 1~V. The ratio of the smallest single-pair sample standard deviation to that of the proposed method ranges from 2.57 to 19.56 at 500~mV and from 2.12 to $2.90\times10^3$ at 1~V. The upper ratio reflects unstable single-pair phase-slope delay estimates for periodic disturbances in the 1-V, nominal 50-km group, where the proposed covariance-weighted fusion maintains meter-scale localization repeatability.
\end{abstract}


\section{Introduction}\label{Intro}
\IEEEPARstart{D}{istributed} optical-fiber sensing uses the fiber itself as an extended sensing element, enabling disturbances to be detected and localized without installing discrete sensors at every measurement point. This capability supports spatially distributed monitoring along the fiber~\cite{Bao2017} and extends geophysical observations to seafloor faults, earthquakes, and ocean dynamics through existing telecommunications cables~\cite{Lindsey2019,Sladen2019}. For vibration sensing, two relevant approaches are backscatter-based interrogation and forward-transmission interferometric interrogation. Optical time-domain reflectometry (OTDR) maps the return time to fiber position, and its phase-sensitive implementations recover local perturbations from the Rayleigh-backscattered field~\cite{Huang2026CPOFDM,Huang2026DMT}. Forward-transmission sensing instead measures perturbation-induced changes in the transmitted optical field~\cite{Marra2018}. When the optical paths follow the permitted amplification directions, inline erbium-doped fiber amplifiers (EDFAs) can compensate for propagation loss and extend the sensing range~\cite{Yan2021}.

The resulting reach is attractive for monitoring long terrestrial and submarine links. Marra \emph{et al.} demonstrated earthquake detection through laser interferometry on terrestrial and submarine cables~\cite{Marra2018}. Mazur \emph{et al.} extracted phase and polarization measurements from a real-time coherent transceiver after 12\,800~km of round-trip transmission over the 6400-km Dunant submarine cable~\cite{Mazur2022}. Such long-range detection does not, by itself, identify where perturbations occur along the fiber. Spatial information can be obtained from position-dependent relative delays between phase signals or from measurements that isolate the phase changes accumulated over individual cable sections. In addition to earthquake detection, Marra \emph{et al.} demonstrated laboratory localization using the propagation-delay difference between phase signals measured at opposite ends of a fiber link~\cite{Marra2018}. Using differences between repeater-loopback phase measurements, Marra \emph{et al.} subsequently resolved environmental signals from separate cable sections~\cite{Marra2022}. Yan \emph{et al.} correlated two differential phase signals in an EDFA-assisted multispan system, demonstrating a 615-km sensing range with 1230~km of total sensing fiber~\cite{Yan2021}. The localization and section-resolved demonstrations illustrate how spatial information can be obtained through propagation-delay comparisons or phase measurements that distinguish different fiber sections.

For interferometric localization, the optical paths must encode disturbance position in measurable relative delays or phase-response features. Mach--Zehnder, Michelson, and Sagnac configurations provide different ways to obtain these observables. Michelson interferometers compare reflected arm fields. Hong \emph{et al.} used two wavelength-separated Michelson interferometers and localized vibrations by correlating their demodulated phase signals~\cite{Hong2011}. In a Sagnac loop, counter-propagating fields encounter the disturbance at different times, producing a position-dependent phase response. Hoffman and Kuzyk extracted position from spectral nulls in the interferometer output~\cite{Hoffman2004}, while Teng \emph{et al.} demonstrated a broadband-source dual-Sagnac sensor for disturbance localization~\cite{Teng2019}.

Dual-Sagnac and hybrid configurations further combine phase responses or control relative timing errors. Hu \emph{et al.} matched optical path lengths in an asymmetric dual-Sagnac sensor to reduce timing errors~\cite{Hu2021}, while Teng \emph{et al.} used spectral-peak ratios of the demodulated phase signals for localization~\cite{Teng2021}. In a merged Sagnac--Michelson configuration, Spammer \emph{et al.} normalized the Sagnac response using the Michelson phase-change rate, reducing dependence on disturbance amplitude~\cite{Spammer1997}. Song \emph{et al.} formed sums and differences of the two interferometric phases and estimated the delay between the resulting waveforms~\cite{Song2020}.

A Mach--Zehnder interferometer (MZI) compares sensing- and reference-arm fields, and paired MZIs can provide separate outputs for relative-delay localization~\cite{Liang2009}. This separate readout makes transmission MZIs suitable for combining direction- and wavelength-dependent delay observations. Di Luch \emph{et al.} demonstrated direction-dependent delay localization using counter-propagating MZIs formed by two fibers in a deployed metropolitan cable~\cite{DiLuch2021}. Chen \emph{et al.} introduced wavelength-dependent delays through dispersion-induced walk-off between two wavelength channels in a unidirectional MZI~\cite{Chen2014WalkOff}. Their implementation also used inline EDFAs in the unidirectional sensing and reference arms, demonstrating compatibility with multispan amplification. These capabilities motivate an MZI implementation that combines propagation direction and wavelength as complementary sources of position information. Their delay sensitivities differ, however: the walk-off delay change per unit displacement is governed by fiber dispersion and wavelength separation. A small differential group delay per unit length therefore converts a given delay-estimation error into a larger position error.

Practical MZI sensors must also address noise and phase-recovery requirements. Ma \emph{et al.} used an asymmetric dual-MZI design with wavelength-division multiplexing to reduce Rayleigh-backscattering noise~\cite{Ma2016ADMZI}. Chen \emph{et al.} employed two self-interference MZIs in a double-ended configuration to extract vibration-induced phase information while relaxing laser-coherence requirements~\cite{Chen2023OL}. Rao \emph{et al.} demonstrated a 151.5-km single-span sensor without optical amplification using compensated self-interference~\cite{Rao2024JLT}. These designs improve the acquisition of phase signals; localization also depends on how their relative delays are estimated and combined.

For relative-delay measurements, estimation precision depends on the signal-to-noise ratio, signal spectrum, and observation time~\cite{Quazi1981}. In MZI and hybrid MZI--Sagnac sensors, phase-spectrum processing has extended localization to more complex disturbances. Ma \emph{et al.} investigated phase-spectrum estimation for multiple-intrusion localization in a hybrid MZI--Sagnac system~\cite{Ma2024}. Rao \emph{et al.} used phase-spectrum delays to locate simultaneous vibrations with distinct frequencies over a 202.3-km forward-transmission link and identified phase unwrapping as a way to extend the upper frequency limit for periodic-signal localization~\cite{Rao2024OE}. Noise propagation and positioning-error limits in hybrid MZI--Sagnac sensing have also been analyzed~\cite{Jin2025}, and dual-sensor beamforming has been applied to recover a desired disturbance in the presence of interfering sources~\cite{Fang2026}. Generalized correlation weights signal spectra before correlation~\cite{Knapp1976}, while multichannel cross-correlation exploits redundancy across sensor observations~\cite{Benesty2004}.

Even for a single dominant disturbance, a precise local delay estimate need not identify the correct position over the full sensing range~\cite{Ianniello1982}. For individual spectral components, higher frequencies produce larger phase changes for the same delay perturbation, while reducing the delay separation corresponding to one phase cycle. When multiple cycle-related delays fall within the physically admissible interval, individual spectral phases can support competing position estimates. Additional phase responses can reject incompatible candidates, yet distinct positions may remain difficult to distinguish when their predicted wrapped observations have separations that are small relative to the effective delay errors~\cite{Akhlaq2016Wavelengths}. A precise estimate within one cycle branch therefore does not establish that the integer ambiguity is correct~\cite{Hassibi1998Integer}. Moreover, pairwise observations derived from shared signals are generally correlated, and their covariance should be accounted for when the observations are combined~\cite{Tucker2023MixedInteger}. The challenge is to combine complementary delay information for reliable ambiguity resolution while retaining fine phase sensitivity.

In this work, we develop covariance-weighted spectral delay fusion with a one-dimensional affine model using a dual-wavelength bidirectional MZI for localization of a single dominant disturbance. The two wavelengths probe the sensing fiber in both directions, providing four phase traces whose six pairwise delays have different position sensitivities. We combine unwrapped delays from cross-spectral phase slopes with wrapped delays from polarity-invariant phase alignment in a common-position fit that accounts for their joint covariance and explicitly estimates the integer ambiguities.
The main contributions are summarized as follows:
\begin{itemize}[topsep=0pt]
\item We present a distributed optical-fiber sensing system using a dual-wavelength bidirectional MZI, in which both wavelengths probe the same sensing fiber from both ends and share a reference branch. Each sensing field interferes with its corresponding reference field, and the two propagation directions use different heterodyne frequencies. We derive the recovered phase traces from the optical fields through photodetection and digital demodulation, establishing how a common disturbance appears in the four channels through wavelength- and direction-dependent delays, with channel-dependent amplitude and polarity. The observation model also retains residual laser phase noise, electrical noise, and waveform distortion.

\item We derive all six pairwise delay--position relations. The observations comprise two same-direction inter-wavelength pairs, two same-wavelength counter-propagating pairs, and two inter-wavelength counter-propagating pairs. The model includes relative timing offsets and distinguishes the dispersion-dependent sensitivities of same-direction pairs from the higher sensitivities of counter-propagating pairs, thereby identifying how delay errors propagate into individual position estimates. With the calibrated model coefficients held fixed, the six delay predictions form a one-dimensional affine line segment parameterized by the disturbance position. We identify five independent constraints, comprising three pairwise closure conditions and two additional propagation conditions, that restrict the admissible delay combinations.

\item We develop a joint estimator that combines unwrapped delays from robust cross-spectral phase slopes with wrapped delays from polarity-invariant phase alignment at a reference frequency selected from the measured spectra. The former exploits coherent phase variation across frequency, while the latter retains phase sensitivity to small delay changes. A generalized least-squares objective fits both delay representations to the affine model. The effective joint covariance accounts for shared-channel and cross-representation dependence, including that introduced by using the unwrapped delays for reference-frequency alignment. Position is solved analytically for each candidate integer ambiguity vector, reducing the remaining optimization to coupled integer least squares. We further derive conditions for exact identifiability and show how periodic alternatives can remain difficult to distinguish when their dispersion-induced delay differences are small relative to the measurement errors. This analysis separates fine local precision from reliable ambiguity resolution.

\item We implement a 131.335-km experimental system at 1530 and 1550~nm and evaluate localization of periodic disturbances at five nominal positions from 25 to 125~km. We compare the proposed method with four single-pair estimators using the same 20 records in each of the reported voltage--distance groups. The proposed method yields sample standard deviations of 1.007--1.685~m at 500~mV and 0.449--1.324~m at 1~V. The ratio of the smallest single-pair standard deviation to the proposed value ranges from 2.57 to 19.56 at 500~mV and from 2.12 to $2.90\times10^3$ at 1~V. The upper ratio at 1~V reflects large single-pair excursions in the nominal 50-km group. The best-performing single pair changes across measurement groups, whereas the proposed method gives the lowest standard deviation in all groups, demonstrating meter-scale localization repeatability across the tested positions and drive levels.
\end{itemize}

The remainder of this paper is organized as follows. Section~\ref{Princ} presents the sensing system and derives the six pairwise delay--position relations. Section~\ref{ProposedMultiFusion} develops spectral delay fusion with the one-dimensional affine model and analyzes position identifiability and periodic ambiguity. Section~\ref{ResDis} presents the experimental results and discussion, and Section~\ref{Conclu} concludes the paper.

\emph{Notation:} Bold uppercase and lowercase symbols denote matrices and vectors, respectively; scalar quantities are nonbold. The upright subscript $\mathrm{v}$ identifies vibration-related quantities. The upright superscript $\mathrm{u}$ labels unwrapped phases, unwrapped delays, and the associated estimation costs. The superscripts $(\cdot)^T$ and $(\cdot)^*$ denote transpose and complex conjugation, respectively, and $|\cdot|$ denotes absolute value or complex modulus. 

\section{Distributed Optical-Fiber Sensing System Using a Dual-Wavelength Bidirectional MZI}\label{Princ}
In Section~\ref{SysMod}, we model the dual-wavelength bidirectional MZI configuration and derive expressions for the four phase traces recovered by heterodyne detection and digital demodulation.
In Section~\ref{DelayModel}, we derive the six pairwise delay--position relations and their position sensitivities, accounting for relative timing offsets.

\subsection{System Model}\label{SysMod}

\begin{figure}[!t]
    \centering
    \includegraphics[width=\columnwidth]{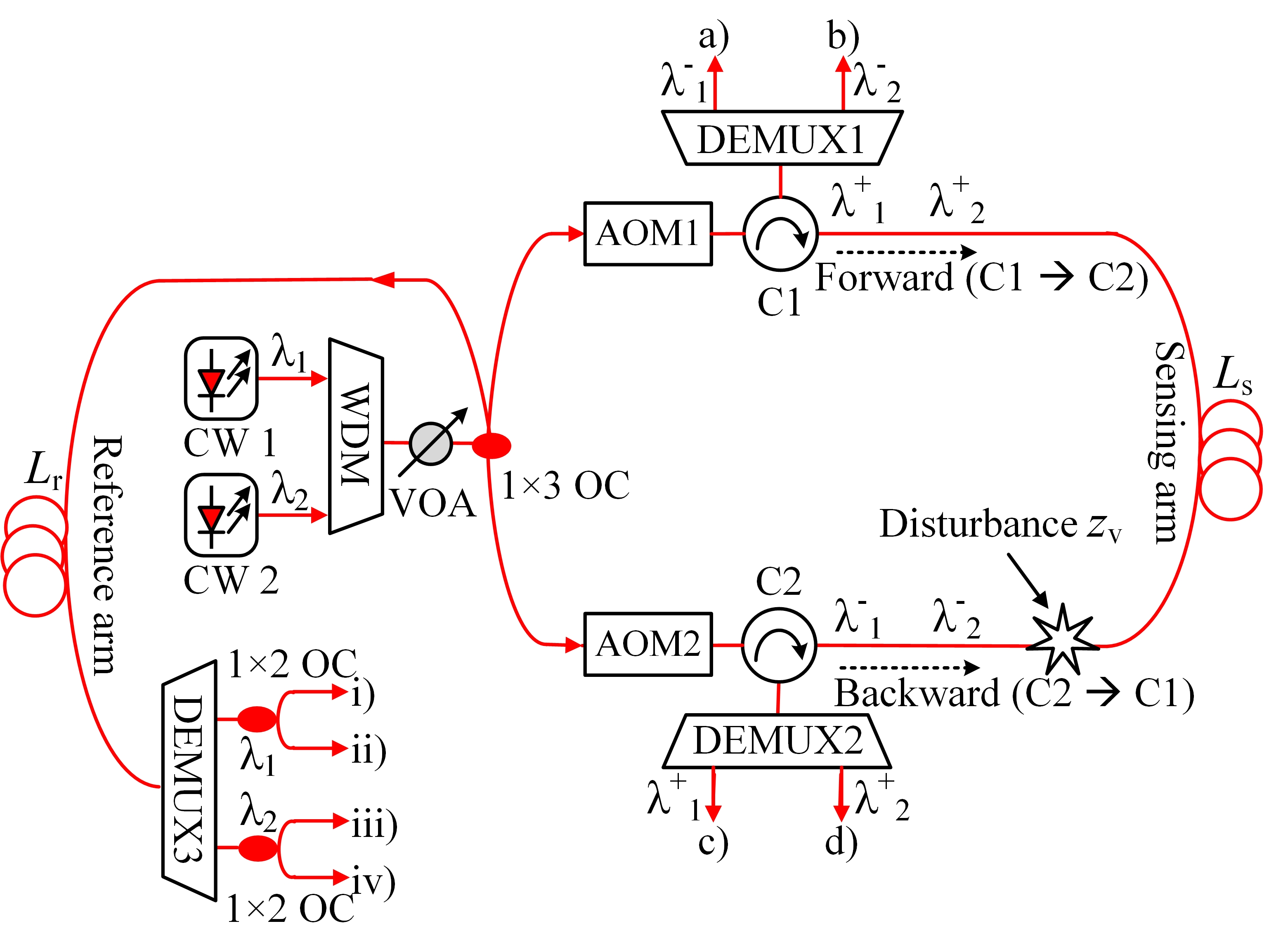}
    \caption{Dual-wavelength bidirectional Mach--Zehnder interferometer for forward-transmission fiber sensing. Receiver connections are omitted; the sensing--reference port pairs are (a,i), (b,iii), (c,ii), and (d,iv). CW: continuous-wave laser; OC: optical coupler; DEMUX: demultiplexer; WDM: wavelength-division multiplexer.}
    \label{fig1}
\end{figure}

Fig.~\ref{fig1} illustrates the dual-wavelength bidirectional MZI configuration of the proposed distributed optical-fiber sensing system. 
Two continuous-wave (CW) carriers at wavelengths $\lambda_1$ and $\lambda_2$ are combined by a wavelength-division multiplexer (WDM) and divided by a $1\times3$ optical coupler (OC) into one reference branch and two sensing branches launched into the same fiber from opposite ends.
The superscripts $(\cdot)^+$ and $(\cdot)^-$ label forward sensing propagation from C1 to C2 and backward sensing propagation from C2 to C1, respectively.
The acousto-optic modulators (AOMs), AOM1 and AOM2, apply the signed frequency offsets $f^+_{\mathrm{A}}$ and $f^-_{\mathrm{A}}$, respectively. Each AOM-shifted sensing field produces a radio-frequency (RF) beat note with its same-wavelength reference field. The two propagation directions are assigned different heterodyne frequencies.

\looseness=1
The sensing fields at ports a)--d) in Fig.~\ref{fig1} are denoted by $E_{\mathrm{s},1}^{-}$, $E_{\mathrm{s},2}^{-}$, $E_{\mathrm{s},1}^{+}$, and $E_{\mathrm{s},2}^{+}$, respectively.
After the demultiplexer (DEMUX) labeled DEMUX3 and the two $1\times2$ OCs, the corresponding reference fields at ports i)--iv) are denoted by $E_{\mathrm{r},1}^{-}$, $E_{\mathrm{r},1}^{+}$, $E_{\mathrm{r},2}^{-}$, and $E_{\mathrm{r},2}^{+}$. For a reference field, the superscript identifies its paired sensing channel, rather than a propagation direction in the reference fiber.
In each receiving channel, the sensing and reference fields at the same wavelength are optically combined before photodetection. The four sensing--reference port pairs are (a,i), (b,iii), (c,ii), and (d,iv), yielding $\{E_{\mathrm{r},m}^{q},E_{\mathrm{s},m}^{q}\}$ for $m\in\{1,2\}$ and $q\in\{+,-\}$.

The four recovered phase traces provide three types of pairwise delay observations: same-wavelength counter-propagating delays, same-direction inter-wavelength delays, and inter-wavelength counter-propagating delays. The two counter-propagating pair types encode the bidirectional propagation geometry, whereas the same-direction inter-wavelength pairs depend on chromatic dispersion. All six observations are obtained from the same four phase traces and are therefore not independent measurement channels.

The optical field emitted by the $m$-th CW laser ($m\in\{1,2\}$) is modeled as
\begin{equation}
    E_{\mathrm{in},m}(t)
    =
    \sqrt{P_{0,m}} e^{j\left(2\pi f_m t+\phi_m(t)\right)}, 
    \label{eq:Ein}
\end{equation}
where $P_{0,m}$, $f_m$, and $\phi_m(t)$ denote the launched optical power, optical carrier frequency, and laser phase noise, respectively.

To keep the model compact while retaining sufficient generality, the insertion losses, splitting ratios, polarization rotations, connector losses, WDM and demultiplexer responses, circulator responses, pigtail responses, and slowly varying intensity fluctuations are represented by equivalent branch responses immediately before photodetection. 
For the reference carrier paired with the $(m,q)$ sensing branch, where $q\in\{+,-\}$, the optical field at the photodiode (PD) input is written as
\begin{equation}
    E_{\mathrm{r},m}^{q}(t)
    =
    h_{\mathrm{r},m}^{q}(t)\sqrt{P_{0,m}} 
    e^{j\left(2\pi f_m t+
    \phi_m(t-\tau_{\mathrm{r},m}^{q})\right)},
    \label{eq:Eref}
\end{equation}
where $\tau_{\mathrm{r},m}^{q}$ is the group delay of the corresponding reference path and $h_{\mathrm{r},m}^{q}(t)$ is the equivalent complex field response of that reference branch.

Similarly, the output AOM-shifted sensing field is modeled as
\begin{equation}
    \begin{aligned}
    E_{\mathrm{s},m}^{q}(t)
    &=h_{\mathrm{s},m}^{q}(t)\sqrt{P_{0,m}}\\
    &\quad\times\exp\!\left\{j\left[
    2\pi(f_m+f_{\mathrm{A}}^q)t
    +\phi_m(t-\tau_{\mathrm{s},m}^{q})\right.\right.\\
    &\hspace{29mm}\left.\left.
    +\varphi_{\mathrm{v},m}^{q}(t-\tau_m^{q})\right]\right\},
    \end{aligned}
    \label{eq:Esens}
\end{equation}
where $\tau_{\mathrm{s},m}^{q}$ is the group delay of the sensing
path, $h_{\mathrm{s},m}^{q}(t)$ is the equivalent complex field
response of the sensing branch, and
$\varphi_{\mathrm{v},m}^{q}\!(\cdot)$ denotes the
disturbance-induced phase modulation observed at wavelength
$\lambda_m$ and direction $q$.
Furthermore, the equivalent complex field response is decomposed as
\begin{equation}
    h_{\mu,m}^{q}(t)
    =
    a_{\mu,m}^{q}(t)e^{j\theta_{\mu,m}^{q}\!(t)},
    \label{eq:h_decomposition}
\end{equation}
where $\mu\in\{\mathrm{r},\mathrm{s}\}$, $a_{\mu,m}^{q}(t)\geq0$ denotes the effective amplitude response and $\theta_{\mu,m}^{q}(t)$ denotes the static and slowly varying phase response. Thus, deterministic phase shifts introduced by the $1\times3$ OC, AOMs, WDMs, demultiplexers, circulators, couplers, pigtails, and connectors are absorbed into $\theta_{\mu,m}^{q}(t)$ rather than expanded separately.

Let $z_{\mathrm{v}}$ be the disturbance position measured from C1 and let $L_{\mathrm{s}}$ be the sensing-fiber length. Within each observation window, the position is taken to be fixed, and the group delay per unit length at each wavelength is modeled as constant along the sensing fiber. Denoting the group velocity at $\lambda_m$ by $v_{\mathrm{g},m}$ and defining $\gamma_m=1/v_{\mathrm{g},m}$, the disturbance-to-receiver delays are
\begin{alignat}{1}
    \tau_m^{+}
    &=
    \gamma_m(L_{\mathrm{s}}-z_{\mathrm{v}}), \label{eq:disturbance_delay1}\\
    \tau_m^{-}
    &=
    \gamma_m z_{\mathrm{v}} .
    \label{eq:disturbance_delay}
\end{alignat}
These direction- and wavelength-dependent delays provide the physical basis for disturbance localization.

For single-port detection in the $(m,q)$ channel, the PD output current is given by
\begin{align}
    i_m^{q}(t)
    &=
    \mathcal{R}_m^{q}
    \left|
    E_{\mathrm{r},m}^{q}(t)+E_{\mathrm{s},m}^{q}(t)
    \right|^2
    +n_m^{q}(t)
    \nonumber\\
    &=
    I_{\mathrm{dc},m}^{q}(t)
    +
    A_m^{q}(t)
    \cos\!\left(
    2\pi f_{\!\mathrm{A}}^{q}t+\psi_m^{q}(t)
    \right)
    +n_m^{q}(t),
    \label{eq:PD_current}
\end{align}
where $\mathcal{R}_m^{q}$ is the responsivity of 
PD and $n_m^{q}(t)$ is the equivalent electrical noise. In~\eqref{eq:PD_current}, the
direct-detection term is $I_{\mathrm{dc},m}^{q}(t)=\mathcal{R}_m^{q}P_{0,m}\left(|h_{\mathrm{r},m}^{q}(t)|^2+|h_{\mathrm{s},m}^{q}(t)|^2\right)$ and the beat amplitude is $A_m^{q}(t)=2\mathcal{R}_m^{q}P_{0,m}|h_{\mathrm{r},m}^{q}(t)h_{\mathrm{s},m}^{q}(t)|$, where the amplitude $A_m^{q}(t)$ includes the effects of optical loss, power fluctuation, polarization mismatch, fading, and RF-chain gain variation. 

The localization information is carried by the interferometric phase in \eqref{eq:PD_current}:
\begin{equation}
    \psi_m^{q}(t)
    =
    \varphi_{\mathrm{v},m}^{q}(t-\tau_m^{q})
    +
    \Delta\phi_m^{q}(t)
    +
    \Theta_m^{q}(t),
    \label{eq:total_phase}
\end{equation}
where $\Delta\phi_m^{q}(t)=\phi_m(t-\tau_{\mathrm{s},m}^{q})-\phi_m(t-\tau_{\mathrm{r},m}^{q})$
is the residual laser phase noise due to the reference--sensing path delay mismatch, and $ \Theta_m^{q}(t) = \theta_{\mathrm{s},m}^{q}(t) - \theta_{\mathrm{r},m}^{q}(t)$
is the relative phase offset between the sensing and reference branch responses, including both static and slowly varying components. 

The digital processing starts from mean removal and denoted as $ x_m^{q}(t) = i_m^{q}(t)-\langle i_m^{q}(t)\rangle$, where $\langle\cdot\rangle$ denotes the temporal average over the recorded block. 
For a nonzero RF beat note, a complex baseband signal is formed by digital downconversion and low-pass filtering~\cite{Kikuchi2016} as
\begin{equation}
    y_m^{q}(t)
    =
    2\mathcal{L}_{B_{\mathrm{LP}}}
    \left\{
    x_m^{q}(t)e^{-j2\pi f_{\!\mathrm{A}}^{q}t}
    \right\},
    \label{eq:DDC_LPF}
\end{equation}
where $\mathcal{L}_{B_{\mathrm{LP}}}\{\cdot\}$ denotes a
zero-phase low-pass filtering operator with bandwidth
$B_{\mathrm{LP}}$. Substituting \eqref{eq:PD_current} into
\eqref{eq:DDC_LPF} gives
\begin{equation}
    y_m^{q}(t)
    \simeq
    A_m^{q}(t)e^{j\psi_m^{q}(t)}
    +
    w_m^{q}(t),
    \label{eq:complex_baseband}
\end{equation}
where $w_m^{q}(t)$ denotes the equivalent complex baseband noise. The
factor of two in \eqref{eq:DDC_LPF} compensates the $1/2$ amplitude
factor introduced by real-valued heterodyne mixing.

The phase is then unwrapped in time, which is written as $\widehat{\psi}_m^{q}(t)  = \operatorname{unwrap} \left\{ \arg\left[y_m^{q}(t)\right]\right\}$~\cite{Itoh1982}.
A slow drift removal operator $\mathcal{D}\{\cdot\}$ is then applied to suppress the mean or linear trend, i.e., $\widetilde{\psi}_m^{q}(t)=\mathcal{D}\left\{\widehat{\psi}_m^{q}(t)\right\}$.
Finally, the vibration-band phase trace is obtained by zero-phase band extraction,
\begin{equation}
    \widehat{\varphi}_m^{q}(t)
    =
    \mathcal{B}_{[f_{\mathrm{L}},f_{\mathrm{H}}]}
    \left\{
    \widetilde{\psi}_m^{q}(t)
    \right\},
    \label{eq:phase_bandpass}
\end{equation}
where $\mathcal{B}_{[f_{\mathrm{L}},f_{\mathrm{H}}]}\{\cdot\}$ denotes a zero-phase bandpass operator over the frequency band $[f_{\mathrm{L}},f_{\mathrm{H}}]$; hence, this operation introduces no deterministic causal-filter group delay.

Combining \eqref{eq:total_phase}--\eqref{eq:phase_bandpass}, the final phase trace can be approximated as
\begin{equation}
    \widehat{\varphi}_m^{q}(t)
    \simeq
    \mathcal{B}_{[f_{\mathrm{L}},f_{\mathrm{H}}]}
    \left\{
    \varphi_{\mathrm{v},m}^{q}(t-\tau_m^{q})
    \right\}
    +
    \widetilde{w}_m^{q}(t),
    \label{eq:final_phase_trace}
\end{equation}
where $\widetilde{w}_m^{q}(t)$ includes the residual laser phase noise, electrical noise, and imperfectly removed phase drift within the selected vibration band. The zero-phase operations in
\eqref{eq:DDC_LPF} and \eqref{eq:phase_bandpass} introduce no deterministic filter group delay; finite-window effects and residual channel-dependent distortion remain in the error term.

Within each observation window containing a single dominant disturbance, the four band-limited disturbance responses are assumed to contain a common disturbance component,
\begin{equation}
\mathcal{B}_{[f_{\mathrm{L}},f_{\mathrm{H}}]}
\left\{\varphi_{\mathrm{v},m}^{q}(t)\right\}
=g_m^q \xi(t)+e_{\mathrm{v},m}^{q}(t),
\label{eq:common_delay_component}
\end{equation}
{
where the nonzero real coefficient $g_m^q$ accounts for channel-dependent amplitude and polarity, $\xi(t)$ denotes the common disturbance waveform, and $e_{\mathrm{v},m}^{q}(t)$ represents residual channel-dependent waveform distortion. Together with~\eqref{eq:final_phase_trace}, this assumption enables pairwise delay inference without requiring identical channel amplitudes; arbitrary channel-dependent phase responses remain outside the ideal common-component model.
\par}

The four phase traces used for delay estimation and multichannel localization are therefore $\widehat{\varphi}_1^{+}(t)$, $\widehat{\varphi}_1^{-}(t)$, $\widehat{\varphi}_2^{+}(t)$, and $\widehat{\varphi}_2^{-}(t)$.

\subsection{Pairwise Delay--Position Relations}\label{DelayModel}
From the four recovered phase traces, let $\tau_i^{+}$ and $\tau_i^{-}$ denote the sensing-fiber propagation delays of the disturbance-induced phase perturbation at wavelength index $i\in\{1,2\}$ in the forward and backward branches, respectively.
The dispersion-induced difference in group delay per unit length is
\begin{equation}
\kappa=\gamma_2-\gamma_1
\simeq D\Delta\lambda  ,
\label{eq:kappa_def}
\end{equation}
where $D$ is the fiber chromatic-dispersion coefficient and $\Delta\lambda=\lambda_2-\lambda_1$. Their units are chosen consistently so that $\kappa$ and $\gamma_i$ both have units of time per unit length.
The six unique physical pairwise delays are defined by
\begin{equation}
 d_{ij}^{qu}(z_{\mathrm{v}})=\tau_i^{q}-\tau_j^{u},
\label{eq:dij_def}
\end{equation}
where $q,u\in\{+,-\}$ and the order of the two channels is retained because it determines the delay sign.
Consequently, the vector of six measured delays is modeled as
\begin{equation}
\widehat{\boldsymbol d}^{\mathrm{meas}}
=\boldsymbol d^{\mathrm{phy}}(z_{\mathrm{v}})+\boldsymbol o+\boldsymbol\epsilon_d,
\label{eq:measured_delay_affine}
\end{equation}
where $\boldsymbol o=[o_1,\ldots,o_6]^T$ collects all position-independent relative timing offsets, including channel- and wavelength-dependent contributions, and $\boldsymbol\epsilon_d$ is the residual delay error. Let $\boldsymbol c=[c_1^+,c_1^-,c_2^+,c_2^-]^T$ collect the effective channel-specific timing offsets. Then $\boldsymbol o=\mathbf B\boldsymbol c$, where the incidence matrix $\mathbf B$ follows the ordered channel pairs defined below.

Only relative channel timing is observable, so $\boldsymbol o$ has three independent degrees of freedom and satisfies pairwise closure.
For an unwrapped measured-delay vector, offset correction gives
\begin{equation}
\widetilde{\boldsymbol d}=\widehat{\boldsymbol d}^{\mathrm{meas}}-\boldsymbol o.
\label{eq:calibrated_delay_vector}
\end{equation}
The resulting individual-pair inversions provide the sensitivity interpretation of the joint model developed below. The same-direction inter-wavelength inversions require $\kappa\neq0$.

\textit{1) Same-direction inter-wavelength delays:}
\begin{equation}
\hat z_{\mathrm{v}}^{(\mathrm{W},+)}
=L_{\mathrm{s}}-\frac{\widetilde d_{21}^{++}}{\kappa},
\label{eq:zd_W_plus}
\end{equation}
\begin{equation}
\hat z_{\mathrm{v}}^{(\mathrm{W},-)}
=\frac{\widetilde d_{21}^{--}}{\kappa}.
\label{eq:zd_W_minus}
\end{equation}

\textit{2) Same-wavelength counter-propagating delays:}
\begin{equation}
\hat z_{\mathrm{v}}^{(\mathrm{B},i)}
=\frac{1}{2}\!\left(L_{\mathrm{s}}+\frac{\widetilde d_{ii}^{-+}}{\gamma_i}\right),
\label{eq:zd_B_i}
\end{equation}
where $i\in\{1,2\}$.

\textit{3) Inter-wavelength counter-propagating delays:}
\begin{equation}
\hat z_{\mathrm{v}}^{(\mathrm{C},1)}
=\frac{\gamma_2L_{\mathrm{s}}-\widetilde d_{21}^{+-}}{\gamma_1+\gamma_2},
\label{eq:zd_C_1}
\end{equation}
\begin{equation}
\hat z_{\mathrm{v}}^{(\mathrm{C},2)}
=\frac{\widetilde d_{21}^{-+}+\gamma_1L_{\mathrm{s}}}{\gamma_1+\gamma_2}.
\label{eq:zd_C_2}
\end{equation}

The six inversions are affected differently by noise, waveform distortion, periodic ambiguity, filter-edge transients, and finite-window delay errors.
In particular, the same-direction inter-wavelength delays in \eqref{eq:zd_W_plus} and \eqref{eq:zd_W_minus} have position sensitivity $|\kappa|$. For $|\kappa|\ll\gamma_1+\gamma_2$, their inversions are more weakly conditioned than those of the counter-propagating pairs.
Their bounded ranges and closure relations provide consistency checks on the delay estimates. Their ability to distinguish candidate positions depends on the predicted delay differences relative to the estimation and calibration errors.  

\section{Spectral Delay Fusion With a One-Dimensional Affine Model}\label{ProposedMultiFusion}
In Section~\ref{AffineModel}, we formulate the one-dimensional affine model for the six pairwise delays and derive unwrapped and wrapped delay estimates from the cross-spectra.
In Section~\ref{JointLocalization}, we jointly estimate position and integer ambiguities using the effective joint covariance and analyze position identifiability and periodic ambiguity.

\subsection{One-Dimensional Affine Model and Cross-Spectral Delay Estimation}\label{AffineModel}\label{SpectralDelay}
A single dominant disturbance is assumed within each observation window, with position $z_{\mathrm{v}}\in\mathcal Z=[z_{\min},z_{\max}]\subseteq[0,L_{\mathrm{s}}]$. The recovered phase traces provide two complementary delay observations: an unwrapped delay obtained from the phase variation across frequency and a wrapped delay obtained from phase alignment at a reference frequency. Both delay representations are fitted to a common disturbance position. The propagation parameters $L_{\mathrm{s}}$ and $\gamma_i$ and the calibrated relative timing offsets $\boldsymbol o$ are held fixed during localization. The formulation requires a sufficiently coherent common component over a nonzero frequency span; periodicity alone does not provide this information.

For the $m$-th ordered pair, let $x_m(t)$ and $y_m(t)$ denote its first and second recovered phase traces. The pair order is 
$(x_1,y_1) =(\widehat\varphi_2^+,\widehat\varphi_1^+)$,
$(x_2,y_2) =(\widehat\varphi_2^-,\widehat\varphi_1^-)$  
$(x_3,y_3) =(\widehat\varphi_1^-,\widehat\varphi_1^+)$, 
$(x_4,y_4) =(\widehat\varphi_2^-,\widehat\varphi_2^+)$, 
$(x_5,y_5) =(\widehat\varphi_2^+,\widehat\varphi_1^-)$, 
and $(x_6,y_6) =(\widehat\varphi_2^-,\widehat\varphi_1^+)$,
where the time argument is omitted for brevity. With $\tau_{x_m}$ and $\tau_{y_m}$ denoting the effective arrival times of the recovered phase traces, the signed delay is consistently defined as
\begin{equation}
d_m=\tau_{x_m}-\tau_{y_m}.
\label{eq:signed_delay_convention}
\end{equation}
The ordered physical delay vector is therefore
\begin{equation}
\boldsymbol d^{\mathrm{phy}}
=[d_{21}^{++},d_{21}^{--},d_{11}^{-+},d_{22}^{-+},d_{21}^{+-},d_{21}^{-+}]^T.
\label{eq:d_order_final}
\end{equation}
Substitution of the propagation delays gives the affine model
\begin{equation}
\boldsymbol d^{\mathrm{pred}}(z_{\mathrm{v}})
=\boldsymbol d^{\mathrm{phy}}(z_{\mathrm{v}})+\boldsymbol o
=\boldsymbol a z_{\mathrm{v}}+{\boldsymbol b},
\label{eq:joint_affine_model}
\end{equation}
In~\eqref{eq:joint_affine_model}, 
\begin{equation}
    \boldsymbol a =[-\kappa,\kappa,2\gamma_1,2\gamma_2,-s,s]^T
    \label{eq:affine_coefficient1}
\end{equation}
and 
\begin{equation}
\boldsymbol b =[\kappa L_{\mathrm{s}},0,-\gamma_1L_{\mathrm{s}},-\gamma_2L_{\mathrm{s}},
\gamma_2L_{\mathrm{s}},-\gamma_1L_{\mathrm{s}}]^T+\boldsymbol o,
\label{eq:affine_coefficient2}
\end{equation}
where $s=\gamma_1+\gamma_2$.
Here, $\boldsymbol a$ contains the signed position sensitivities, whereas $\boldsymbol b$ contains the position-independent propagation terms and fixed timing offsets.

With these parameters fixed, all six predicted delays depend on the same disturbance position. Their admissible values form a one-dimensional affine line segment in the six-dimensional delay space, parameterized by $z_{\mathrm{v}}\in\mathcal Z$. Pairwise closure alone leaves three degrees of freedom; the common propagation law leaves only the position unknown. The line direction is determined by the wavelength- and direction-dependent sensitivities in $\boldsymbol a$. For $M\geq2$ retained pairs with nonzero sensitivity vector, the underlying line can equivalently be specified by
\begin{align}
\mathbf C\boldsymbol a &=\boldsymbol 0, \label{eq:affine_constraint1}\\
\mathbf C\boldsymbol d^{\mathrm{pred}} &=\boldsymbol c_{\mathrm{pred}}, \label{eq:affine_constraints}
\end{align}
where $\mathbf C\in\mathbb R^{(M-1)\times M}$ has full row rank and $\boldsymbol c_{\mathrm{pred}} =\mathbf C\boldsymbol b$. Vectors and matrices are restricted to the same pair set when $M$ is smaller than the full set. For the six-pair model, one admissible constraint matrix is
\begin{equation}
\mathbf C=
\begin{bmatrix}
1&1&0&0&0&0\\
0&0&0&0&1&1\\
0&0&-\gamma_2/\gamma_1&1&0&0\\
-1&0&1&0&1&0\\
1&0&\kappa/(2\gamma_1)&0&0&0
\end{bmatrix}.
\label{eq:C_matrix_final}
\end{equation}
The five independent constraints combine three pairwise closure conditions with two additional conditions imposed by the calibrated propagation model. Restricting the model to the four high-sensitivity pairs leaves three independent constraints. In either case, the number of constraints specifies the dimension of the admissible delay set; it does not, by itself, quantify the separation between periodic aliases.
The one-dimensional parameterization is conditional on the fixed model coefficients; the relative timing offsets are not additional free variables in the position search.

The affine model specifies the delay predictions associated with each candidate position. We next obtain the unwrapped and wrapped delay observations to be fitted to these predictions.

Let $X_{m,j}(f_k)$ and $Y_{m,j}(f_k)$ be the Fourier transforms of the ordered phase traces in the $j$-th analysis segment, where $j=1,\ldots,K$ and $f_k$ denotes a sampled frequency. Using segment averaging for spectral estimation~\cite{Welch1967}, define the averaged cross-spectrum and magnitude-squared coherence as
\begin{equation}
\widehat S_{xy,m}(f_k)
=\frac{1}{K}\sum_{j=1}^{K}X_{m,j}(f_k)Y_{m,j}^{*}(f_k),
\label{eq:cross_spectrum}
\end{equation}
\begin{equation}
\chi_m(f_k)
=\frac{|\widehat S_{xy,m}(f_k)|^2}
{\widehat S_{xx,m}(f_k)\widehat S_{yy,m}(f_k)},
\label{eq:spectral_coherence}
\end{equation}
The auto-spectra $\widehat S_{xx,m}$ and $\widehat S_{yy,m}$ are defined by the same segment average as in \eqref{eq:cross_spectrum}, with the cross-products replaced by $|X_{m,j}|^2$ and $|Y_{m,j}|^2$, respectively. Only frequencies with nonzero spectral support are considered. Coherence characterizes the linear dependence between the two phase traces and informs delay-estimation reliability~\cite{Carter1987}. Here, we use coherence and spectral magnitude as indicators of the reliability of the phase observations. These indicators are not evaluated as functions of the candidate position.

Under the common-component assumption in \eqref{eq:common_delay_component}, the cross-spectrum is approximated by
\begin{equation}
\widehat S_{xy,m}(f)
\simeq A_m(f)
\exp\!\left\{j\beta_m-j2\pi f d_m^{\mathrm{pred}}(z_{\mathrm{v}})\right\}
+e_{S,m}(f),
\label{eq:cross_spectral_delay_model}
\end{equation}
where $A_m(f)\geq0$, $\beta_m\in\{0,\pi\}$ represents the relative polarity, and $e_{S,m}(f)$ includes noise, finite-window effects, and residual waveform distortion. The negative phase slope follows directly from the signed-delay convention in \eqref{eq:signed_delay_convention}. A deterministic linear phase response is included in the fixed timing offset. An arbitrary remaining frequency-dependent channel phase is not removed by this delay model. Denote its ideal cross-spectrum, obtained by omitting $e_{S,m}$, by $S_{xy,m}$.

\subsubsection{Unwrapped Delay From Phase Slope}

On a connected frequency interval with reliable phase observations, \eqref{eq:cross_spectral_delay_model} gives
\begin{equation}
\frac{\partial}{\partial f}\arg S_{xy,m}(f)
=-2\pi d_m^{\mathrm{pred}}(z_{\mathrm{v}}),
\label{eq:phase_slope_delay}
\end{equation}
where the phase is interpreted continuously in the ideal common-component limit. Let $\mathcal I_{m,r}$ denote the $r$-th connected reliable frequency interval, and let $\phi_{m,k}^{\mathrm u}$ be the phase unwrapped within that interval. A general robust phase-slope estimate is
\begin{equation}
\begin{aligned}
&(\widehat d_m^{\mathrm u},\{\widehat\beta_{m,r}\})
\in\arg\min_{d,\{\beta_{m,r}\}}
\sum_r\sum_{k\in\mathcal I_{m,r}}v_{m,k}\\[-1mm]
&\hspace{8mm}\times
\rho\!\left(
\frac{\phi_{m,k}^{\mathrm u}-\beta_{m,r}
+2\pi(f_k-f_0)d}{\sigma_{\phi,m}}
\right).
\end{aligned}
\label{eq:unwrapped_delay_estimate}
\end{equation}
Here, $v_{m,k}\geq0$ is a phase-reliability weight, $\rho$ is a robust loss, such as the Huber loss~\cite{Huber1964}, $\sigma_{\phi,m}>0$ is a phase-residual scale, and $f_0>0$ is the reference frequency. The interval-specific intercepts accommodate independent phase-unwrapping constants without imposing continuity across unreliable spectral gaps. They are nuisance parameters for slope estimation, not estimates used to remove arbitrary phase offsets from the wrapped observation below.

For uniformly spaced frequencies with spacing $\Delta f$, phase differences identify a delay modulo $1/\Delta f$. Thus, the unwrapped delay refers to the branch selected within a prescribed delay interval, not to an unlimited ambiguity-free measurement. A sufficient ideal condition for using the principal phase-gradient branch is
\begin{equation}
\left|d_m^{\mathrm{pred}}(z_{\mathrm{v}})\right|
<\frac{1}{2\Delta f}, 
\label{eq:spectral_delay_interval}
\end{equation}
where $z_{\mathrm{v}}\in\mathcal Z$.
At least one connected reliable frequency interval must have a nonzero frequency span. A single spectral line cannot determine a slope, and leakage caused by windowing that line is not independent delay information.

\subsubsection{Wrapped Delay From Reference-Frequency Phase}

The unwrapped delay is used to align the phase observations to the common reference frequency:
\begin{equation}
\theta_{m,k}
=\arg\widehat S_{xy,m}(f_k)
+2\pi(f_k-f_0)\widehat d_m^{\mathrm u}.
\label{eq:reference_frequency_alignment}
\end{equation}
To remove the unknown real-valued polarity, define the doubled-phase average
\begin{equation}
q_m=
\frac{\sum_k \omega_{m,k}\exp(j2\theta_{m,k})}
{\sum_k\omega_{m,k}},
\label{eq:doubled_phase_average}
\end{equation}
where $\omega_{m,k}\geq0$ weights reliable spectral observations and $\sum_k\omega_{m,k}>0$. For $q_m\neq0$, the resulting wrapped delay is
\begin{equation}
\widehat r_m
=\mathcal W_{T_{\mathrm a}}\!\left(
-\frac{\arg q_m}{4\pi f_0}
\right),
\label{eq:spectral_wrapped_delay}
\end{equation}
with $\mathcal W_T(x)=\mathrm{mod}(x+T/2,T)-T/2$ and $T_{\mathrm a}={1}/(2f_0)$. In the ideal limit $\widehat d_m^{\mathrm u}=d_m^{\mathrm{pred}}$, the aligned doubled phase is $-4\pi f_0d_m^{\mathrm{pred}}$ modulo $2\pi$, giving \eqref{eq:spectral_wrapped_delay}. The ambiguity period is therefore determined by the polarity-invariant phase representation. The period need not equal a burst-repetition interval.

Equation~\eqref{eq:doubled_phase_average} removes a phase reversal of $\pi$, but not an arbitrary channel phase offset. If a residual constant phase $\beta_m$ is not an integer multiple of $\pi$, the ideal wrapped delay becomes $\mathcal W_{T_{\mathrm a}}(d_m^{\mathrm{pred}}-\beta_m/(2\pi f_0))$. The real-gain common-component assumption is consequently required by the wrapped-delay model, even though a free intercept makes the unwrapped slope insensitive to a constant phase.

The wrapped observation uses $\widehat d_m^{\mathrm u}$ in \eqref{eq:reference_frequency_alignment} and is derived from the same phase traces. It is therefore correlated with the unwrapped observation. A value of $|q_m|$ close to unity indicates consistent phase alignment within the adopted model; it does not establish the correct integer ambiguity.

\subsection{Joint Localization and Ambiguity Analysis}\label{JointLocalization}

The two delay representations and the affine model define a problem of jointly estimating position and integer ambiguities. We formulate the estimator and then examine whether the available observations distinguish competing positions.

Let $\widehat{\boldsymbol d}^{\mathrm u}$ and $\widehat{\boldsymbol r}$ collect the unwrapped and wrapped delay estimates for the $M$ retained pairs. Both represent the same predicted delay vector $\boldsymbol d^{\mathrm{pred}}(z_{\mathrm{v}})=\boldsymbol a z_{\mathrm{v}}+\boldsymbol b$, but the latter is observed modulo $T_{\mathrm a}$. Their joint observation model is
\begin{equation}
\begin{aligned}
\widehat{\boldsymbol d}^{\mathrm u}
&=\boldsymbol a z_{\mathrm{v}}+\boldsymbol b+\boldsymbol\epsilon_{\mathrm u},\\
\widehat{\boldsymbol r}
&=\mathcal W_{T_{\mathrm a}}\!\left(
\boldsymbol a z_{\mathrm{v}}+\boldsymbol b+\boldsymbol\epsilon_{\mathrm r}
\right),
\end{aligned}
\label{eq:joint_observation_model}
\end{equation}
where $\boldsymbol\epsilon_{\mathrm u}$ and $\boldsymbol\epsilon_{\mathrm r}$ denote the residual delay errors before wrapping, and $\mathcal W_{T_{\mathrm a}}$ acts elementwise.

Let $\boldsymbol n\in\mathbb Z^M$ denote the integer ambiguity vector. Each component specifies an integer multiple of the wrapping period $T_{\mathrm a}=1/(2f_0)$. A candidate vector gives the restored delays
\begin{equation}
\boldsymbol d_{\boldsymbol n}
=\widehat{\boldsymbol r}+T_{\mathrm a}\boldsymbol n.
\label{eq:integer_restored_delay}
\end{equation}
For a given $\boldsymbol n$, the two delay representations are expressed on the same unwrapped delay scale and form the augmented observation
\begin{equation}
\boldsymbol y_{\boldsymbol n}
=\begin{bmatrix}
\widehat{\boldsymbol d}^{\mathrm u}\\
\boldsymbol d_{\boldsymbol n}
\end{bmatrix}
\in\mathbb R^{2M}.
\label{eq:augmented_delay_observation}
\end{equation}
Thus, $\boldsymbol y_{\boldsymbol n}$ is determined by the measured delays and the candidate integer ambiguity vector.

The two blocks of $\boldsymbol y_{\boldsymbol n}$ estimate the same ordered pairwise delays. Consequently, their physical prediction at a candidate position repeats $\boldsymbol d^{\mathrm{pred}}(z_{\mathrm{v}})$ in both blocks:
\begin{equation}
\boldsymbol\mu(z_{\mathrm{v}})
=\begin{bmatrix}
\boldsymbol d^{\mathrm{pred}}(z_{\mathrm{v}})\\
\boldsymbol d^{\mathrm{pred}}(z_{\mathrm{v}})
\end{bmatrix}
=\boldsymbol h z_{\mathrm{v}}+\boldsymbol g,
\label{eq:augmented_delay_model}
\end{equation}
where $\boldsymbol h=[\boldsymbol a^T,\boldsymbol a^T]^T$ and $\boldsymbol g=[\boldsymbol b^T,\boldsymbol b^T]^T$ are fixed augmented model coefficients. The difference $\boldsymbol y_{\boldsymbol n}-\boldsymbol\mu(z_{\mathrm{v}})$ therefore measures how well a candidate position and integer ambiguity vector jointly explain the two delay observations.

The residual errors are generally correlated because the pairwise estimates share the same phase traces and the wrapped-delay construction uses the unwrapped-delay estimate for reference-frequency alignment. To account for these dependencies, we introduce the effective joint delay covariance
\begin{equation}
\boldsymbol\Sigma=
\begin{bmatrix}
\boldsymbol\Sigma_{\mathrm{uu}}&
\boldsymbol\Sigma_{\mathrm{ur}}\\
\boldsymbol\Sigma_{\mathrm{ur}}^T&
\boldsymbol\Sigma_{\mathrm{rr}}
\end{bmatrix}\succ\boldsymbol 0,
\label{eq:joint_delay_covariance}
\end{equation}
where $\boldsymbol\Sigma\in\mathbb R^{2M\times2M}$. The diagonal blocks describe errors within each delay representation, including correlations among pairs that share a channel. The off-diagonal block $\boldsymbol\Sigma_{\mathrm{ur}}$ accounts for dependence between the two representations. The wrapped-delay errors are represented locally for a fixed integer ambiguity vector. Alternative integer vectors are considered explicitly through $\boldsymbol n$, without inflating the covariance to represent their separation. The matrix is held fixed during each position search and may be regularized to account for residual model mismatch. Unless independently validated as a sampling covariance, it serves as an effective error-weighting model rather than a calibrated measure of localization confidence.

The position and integer ambiguity vector are jointly estimated by generalized least squares:
\begin{align}
Q(z_{\mathrm{v}},\boldsymbol n)
&=\boldsymbol e_{\boldsymbol n}(z_{\mathrm{v}})^T
\boldsymbol\Sigma^{-1}
\boldsymbol e_{\boldsymbol n}(z_{\mathrm{v}}), \label{eq:joint_integer_costadd} \\
\boldsymbol e_{\boldsymbol n}(z_{\mathrm{v}})
&=\boldsymbol y_{\boldsymbol n}-\boldsymbol h z_{\mathrm{v}}-\boldsymbol g, \label{eq:joint_integer_cost}
\end{align}
\begin{equation}
(\widehat z_{\mathrm{v}},\widehat{\boldsymbol n}) = \arg\min_{\substack{z_{\mathrm{v}}\in\mathcal Z\\
\boldsymbol n\in\mathbb Z^M}}
Q(z_{\mathrm{v}},\boldsymbol n).
\label{eq:joint_integer_position_estimate}
\end{equation}
Thus, the fitted augmented delays lie on the one-dimensional affine line defined by $\boldsymbol\mu(z_{\mathrm{v}})$, while the observations may deviate from that line according to the adopted error model. A separate physical-constraint penalty is not required. In particular, a restored delay vector with an inconsistent combination of cycle indices cannot be fitted exactly by any common position. A coherent shift to a nearby alias can nevertheless have a small residual, as analyzed below.

The role of the unwrapped observation can be made explicit by defining
\begin{align}
\mathbf K&=\boldsymbol\Sigma_{\mathrm{ur}}^T
\boldsymbol\Sigma_{\mathrm{uu}}^{-1},\label{eq:conditional_covarianceadd}\\
\mathbf S&=\boldsymbol\Sigma_{\mathrm{rr}}
-\boldsymbol\Sigma_{\mathrm{ur}}^T
\boldsymbol\Sigma_{\mathrm{uu}}^{-1}
\boldsymbol\Sigma_{\mathrm{ur}} \label{eq:conditional_covariance} .
\end{align}
With $\boldsymbol e_{\mathrm u}=\widehat{\boldsymbol d}^{\mathrm u}-\boldsymbol a z_{\mathrm{v}}-\boldsymbol b$ and $\boldsymbol e_{\mathrm r}=\boldsymbol d_{\boldsymbol n}-\boldsymbol a z_{\mathrm{v}}-\boldsymbol b$, the joint cost becomes
\begin{equation}
Q
=\boldsymbol e_{\mathrm u}^T\boldsymbol\Sigma_{\mathrm{uu}}^{-1}
\boldsymbol e_{\mathrm u} +(\boldsymbol e_{\mathrm r}-\mathbf K\boldsymbol e_{\mathrm u})^T
\mathbf S^{-1}
(\boldsymbol e_{\mathrm r}-\mathbf K\boldsymbol e_{\mathrm u}).
\label{eq:conditional_cost_decomposition}
\end{equation}
The first term penalizes candidate positions that differ from the unwrapped delay estimates. The second term measures the additional agreement of the restored, wrapped delays, accounting for their correlation with the unwrapped errors. This decomposition prevents the two delay representations from being treated as independent evidence. Same-direction inter-wavelength pairs enter this joint objective through their position sensitivities and error covariance without an additional position-dependent score.

For a fixed integer vector, the cost in \eqref{eq:joint_integer_costadd}, with the residual defined in \eqref{eq:joint_integer_cost}, is quadratic in position. Its constrained minimizer is
\begin{equation}
z^{\star}(\boldsymbol n)
=\Pi_{\mathcal Z}\!\left[
\frac{\boldsymbol h^T\boldsymbol\Sigma^{-1}
(\boldsymbol y_{\boldsymbol n}-\boldsymbol g)}
{\boldsymbol h^T\boldsymbol\Sigma^{-1}\boldsymbol h}
\right],
\label{eq:conditional_position_solution}
\end{equation}
where $\Pi_{\mathcal Z}$ denotes projection onto the admissible interval. Substituting \eqref{eq:conditional_position_solution} into the joint cost reduces the remaining optimization to a coupled integer least-squares problem~\cite{Teunissen1995}. The coupling generally prevents the integer ambiguities from being determined by independent pairwise rounding.

Equivalently, define the position-domain profile cost and relative score
\begin{equation}
Q_{\mathrm p}(z_{\mathrm{v}})
=\min_{\boldsymbol n\in\mathbb Z^M}Q(z_{\mathrm{v}},\boldsymbol n),
\label{eq:profile_position_cost}
\end{equation}
\begin{equation}
\mathcal L_{\mathrm{rel}}(z_{\mathrm{v}})
=\exp\!\left\{-\frac{
Q_{\mathrm p}(z_{\mathrm{v}})
-\min_{\zeta\in\mathcal Z}Q_{\mathrm p}(\zeta)}{2}\right\}.
\label{eq:relative_position_score}
\end{equation}
The location at the maximum of $\mathcal L_{\mathrm{rel}}$, or equivalently the minimum of $Q_{\mathrm p}$, is reported directly as $\widehat z_{\mathrm{v}}$. A fixed convention resolves exact ties. Under a working Gaussian error model, $\exp[-Q(z_{\mathrm{v}},\boldsymbol n)/2]$ is proportional to a Gaussian likelihood term for a specified integer ambiguity vector. Equation~\eqref{eq:relative_position_score} maximizes this term over the integers and normalizes its maximum to unity; it is not the marginal likelihood of the wrapped observations, which would sum the contributions over $\boldsymbol n$. The adopted relative score represents the generalized least-squares objective, not a probability of correct localization.

The one-dimensional delay model restricts the admissible delay combinations, but distinguishing candidate integer ambiguities also depends on the position sensitivities and the available delay information. In the noiseless wrapped model, two positions separated by $\Delta z$ are indistinguishable precisely when
\begin{equation}
\boldsymbol a\,\Delta z=T_{\mathrm a}\boldsymbol k,\qquad\boldsymbol k\in\mathbb Z^M,
\label{eq:exact_wrapped_alias}
\end{equation}
with both positions in $\mathcal Z$. Fixed timing offsets cancel from this condition.

For the high-sensitivity pairs $\{d_3,d_4,d_5,d_6\}$, the sensitivity vector is $[s-\kappa,s+\kappa,-s,s]^T$. A displacement $\Delta z_{\mathrm a}=T_{\mathrm a}/s$ produces
\begin{equation}
\boldsymbol a_{\mathrm H}\Delta z_{\mathrm a}
=T_{\mathrm a}\begin{bmatrix}1\\1\\-1\\1\end{bmatrix}
+\begin{bmatrix}
-\kappa\Delta z_{\mathrm a}\\
\kappa\Delta z_{\mathrm a}\\
0\\0
\end{bmatrix}.
\label{eq:near_common_position_alias}
\end{equation}
Thus, the integer multiples of $T_{\mathrm a}$ vanish under wrapping, leaving only the delay differences proportional to $\kappa$. When these differences are small relative to the effective delay errors, adjacent branches can remain difficult to distinguish even though the constraint matrix has the required rank.

Exact identifiability must be distinguished from the ability to discriminate positions in the presence of measurement errors. For example, if $\kappa\neq0$ and
\begin{equation}
2|\kappa|(z_{\max}-z_{\min})<T_{\mathrm a},
\label{eq:sufficient_wrapped_identifiability}
\end{equation}
then the high-sensitivity observations have no exact nonzero alias within $\mathcal Z$. Indeed, the difference between the $d_4$ and $d_3$ components of \eqref{eq:exact_wrapped_alias} requires $2\kappa\Delta z=T_{\mathrm a}(k_4-k_3)$; \eqref{eq:sufficient_wrapped_identifiability} forces $k_4=k_3$ and hence $\Delta z=0$. This noiseless result does not imply that the remaining separation between the predicted wrapped delays can be resolved in a noisy record.

The unwrapped phase slope provides additional information for distinguishing candidate positions through the coherent frequency dependence of the measured waveform. A finite-duration periodic disturbance can provide such information through its resolved spectral structure, subject to the common-component model and sufficient phase reliability. This is a measurement requirement, not information generated by the physical constraints. The method uses this measured phase variation across frequency without adding a separate position-dependent score based on correlation magnitude.

Finally, with the integer ambiguity vector held fixed, the cost $Q(z_{\mathrm{v}},\widehat{\boldsymbol n})$ has positive curvature $2\boldsymbol h^T\boldsymbol\Sigma^{-1}\boldsymbol h$ regardless of whether $\widehat{\boldsymbol n}$ is correct. Its local width characterizes conditional position sensitivity under the adopted error model, not global ambiguity resolution. Alternative integer ambiguity vectors are considered in the profile cost defined by \eqref{eq:profile_position_cost}. Localization repeatability and agreement with an independently known position therefore address different aspects of localization performance.

\section{Experimental Results and Discussion}\label{ResDis}
\begin{figure}[!t]
    \centering
    \includegraphics[width=0.48\textwidth]{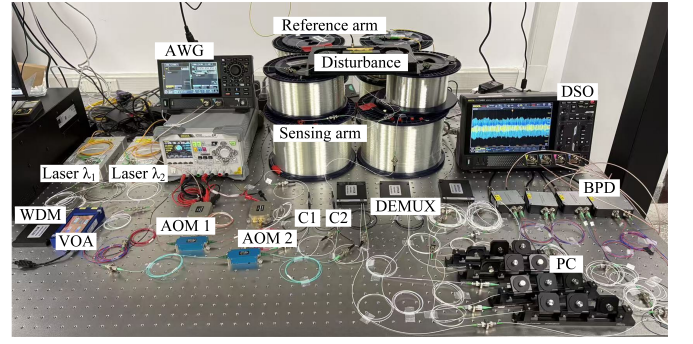}
    \caption{
    Experimental setup of the dual-wavelength bidirectional Mach--Zehnder interferometer for high-precision forward-transmission fiber sensing. AWG: arbitrary waveform generator; WDM: wavelength-division multiplexer; VOA: variable optical attenuator; AOM: acousto-optic modulator; PC: polarization controller; DEMUX: demultiplexer; BPD: balanced photodetector; DSO: digital storage oscilloscope.}
    \label{fig:Exp}
\end{figure}

The experimental setup in Fig.~\ref{fig:Exp} implements the dual-wavelength bidirectional MZI shown in Fig.~\ref{fig1}. Two CW lasers (FL-SF-1530-S and FL-SF-1550-S, Precilasers, China) provide the carriers at $\lambda_1=1530$~nm and $\lambda_2=1550$~nm, respectively. Their outputs are combined by a WDM, adjusted by a variable optical attenuator (VOA), and divided by a $1\times3$ OC into one reference branch and two counter-propagating sensing branches. AOM1 (SGTF200-1550-1P) and AOM2 (SGTF80-1550-1), both from Chongqing Smart Science \& Technology, China, provide frequency shifts of 200 and 80~MHz in the forward C1~$\to$~C2 and backward C2~$\to$~C1 sensing branches, respectively.

The sensing and reference arms use YOFC fiber, with lengths $L_{\mathrm{s}}\simeq131.3350$~km and $L_{\mathrm{r}}\simeq131.3973$~km, respectively. The reference fiber is approximately 62~m longer than the sensing fiber. Residual laser phase noise associated with the total sensing--reference path delay mismatch is retained through $\Delta\phi_m^{q}(t)$ in \eqref{eq:total_phase}. A 1550-nm lithium-niobate (LiNbO$_3$) phase modulator (PM) is inserted at position $z_{\mathrm{v}}$ in the sensing arm and driven by an arbitrary waveform generator (AWG, RIGOL DG912 Pro). The PM emulates the optical-phase perturbation induced by a localized vibration, as in~\cite{Chen2014WalkOff}. Measurements are performed at drive-voltage settings of 500~mV and 1~V. The results below focus exclusively on periodic excitations to examine the additional localization challenge posed by integer-cycle ambiguity.

At each end of the sensing arm, a demultiplexer separates the two wavelength channels. After transmission through the shared reference arm, the two wavelength channels are demultiplexed and each is divided by a $1\times2$ OC to supply the reference fields for the two sensing directions. Polarization controllers (PCs) adjust the reference--sensing polarization alignment before interference, and four InGaAs balanced photodetectors (BPDs, including UBD-500M-A) recover the heterodyne signals. Equation~\eqref{eq:PD_current} describes single-port detection; the balanced receivers recover the same reference--sensing beat phase, with receiver-dependent amplitude factors absorbed into $A_m^{q}(t)$. 
The four outputs are recorded together by a digital storage oscilloscope (DSO, RIGOL DHO4404) at 1~GSa/s, with $10^7$ samples per channel in each 10-ms record. Consistent with ports a)--d) in Fig.~\ref{fig1}, the recovered phase traces are $\widehat\varphi_1^{-}$, $\widehat\varphi_2^{-}$, $\widehat\varphi_1^{+}$, and $\widehat\varphi_2^{+}$, respectively.

\begin{figure}[!b]
    \centering
    \includegraphics[width=0.48\textwidth]{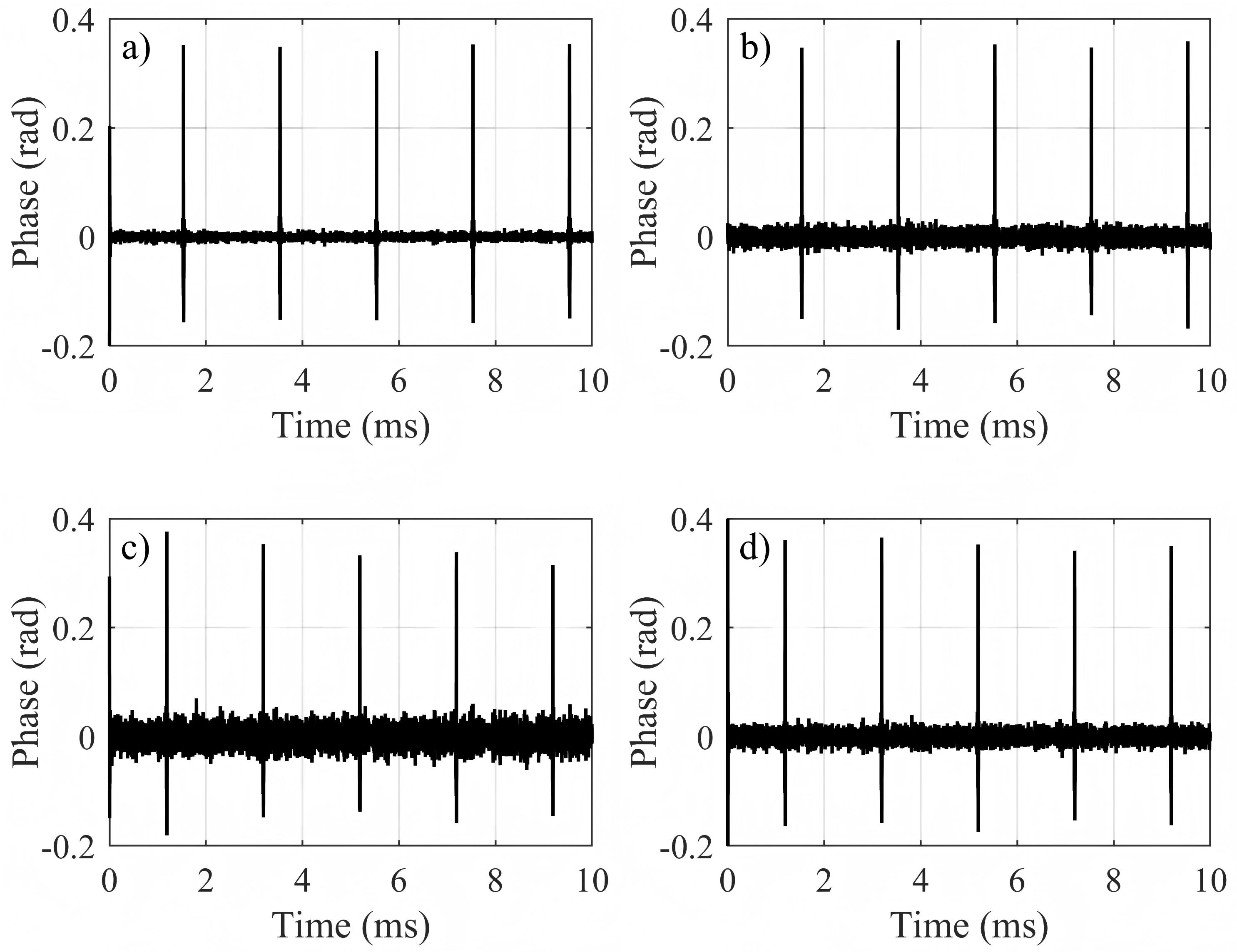}
    \caption{Recovered band-limited phase traces for a 500-mV measurement in the nominal 100-km group: (a) $\widehat\varphi_1^{-}$, (b) $\widehat\varphi_2^{-}$, (c) $\widehat\varphi_1^{+}$, and (d) $\widehat\varphi_2^{+}$.} 
    \label{Resfig:500mvfour}
\end{figure}

Fig.~\ref{Resfig:500mvfour} shows the recovered phase traces for a 500-mV measurement in the nominal 100-km group. The nominal distance labels are based on coarse position measurements obtained with an OTDR instrument and identify the corresponding measurement groups rather than exact disturbance positions. Panels (a)--(d) correspond to the four ports in the order given above. The recorded voltages are digitally downconverted and low-pass filtered at 1~MHz according to \eqref{eq:DDC_LPF}. The phase is then unwrapped and linearly detrended, followed by 60--260-kHz bandpass filtering according to \eqref{eq:phase_bandpass}, after which the traces are decimated to 2~MSa/s. Namely, the plotted signals are band-limited phase observations. All four traces exhibit repeated localized features with different levels of background fluctuations and time displacement between the propagation directions. The latter are particularly visible in the 1530-nm forward trace in Fig.~\ref{Resfig:500mvfour}~(c).

The four traces provide the six ordered pairwise delays in \eqref{eq:d_order_final}. Cross-spectra and spectral coherence are estimated over five nonoverlapping 2-ms blocks using \eqref{eq:cross_spectrum} and \eqref{eq:spectral_coherence}, giving a frequency spacing of 500~Hz. For each pair, the robust phase-slope fit in \eqref{eq:unwrapped_delay_estimate} yields $\widehat d_m^{\mathrm u}$. Reference-frequency alignment and doubled-phase averaging in \eqref{eq:reference_frequency_alignment}--\eqref{eq:spectral_wrapped_delay} yield $\widehat r_m$. The common reference frequency is selected from the reliable measured cross-spectra. For the record considered here, $f_0\simeq121.830$~kHz and $T_{\mathrm a}\simeq4.104~\mu\mathrm{s}$; neither an assumed vibration frequency nor the burst-repetition interval is used to specify this ambiguity period.

All six pairs are retained in the joint estimator. In particular, $d_1=d_{21}^{++}$ and $d_2=d_{21}^{--}$ compare the two wavelengths propagating in the same direction. Their predictions are $d_1^{\mathrm{pred}}=\kappa(L_{\mathrm{s}}-z_{\mathrm{v}})+o_1$ and $d_2^{\mathrm{pred}}=\kappa z_{\mathrm{v}}+o_2$, where $o_1$ and $o_2$ are calibrated relative channel timing offsets that remain fixed during localization. In our experiment with the 20-nm wavelength separation, the position sensitivities $-\kappa$ and $\kappa$ are much smaller in magnitude than those of $d_3$--$d_6$. Directly dividing their delay errors by $\kappa$ would therefore give poorly conditioned individual position estimates. Instead, both their unwrapped and wrapped observations enter \eqref{eq:joint_integer_costadd} and \eqref{eq:joint_integer_cost} with $M=6$, using the effective covariance in \eqref{eq:joint_delay_covariance}. For this record, $\widehat n_1=\widehat n_2=0$; the two pairs, i.e., $d_1$ and $d_2$ remain in the fit and are not treated as separate position anchors. 

The joint fit uses fixed propagation coefficients, relative timing offsets, and the effective joint delay covariance in \eqref{eq:joint_delay_covariance} as its error-weighting model. These inputs are specified before the position search. The OTDR measurements provide the spatial references used in the calibration. Under the available experimental conditions, these measurements provide only coarse position information on the scale of several meters to a few tens of meters and cannot verify meter-scale or submeter absolute localization accuracy. The experimental evaluation therefore distinguishes coarse positional agreement with the OTDR references from localization repeatability. Localization repeatability is quantified by the sample standard deviation relative to each method's mean. A small value indicates more repeatable estimates.

\begin{figure}[!t]
    \centering
    \includegraphics[width=0.46\textwidth]{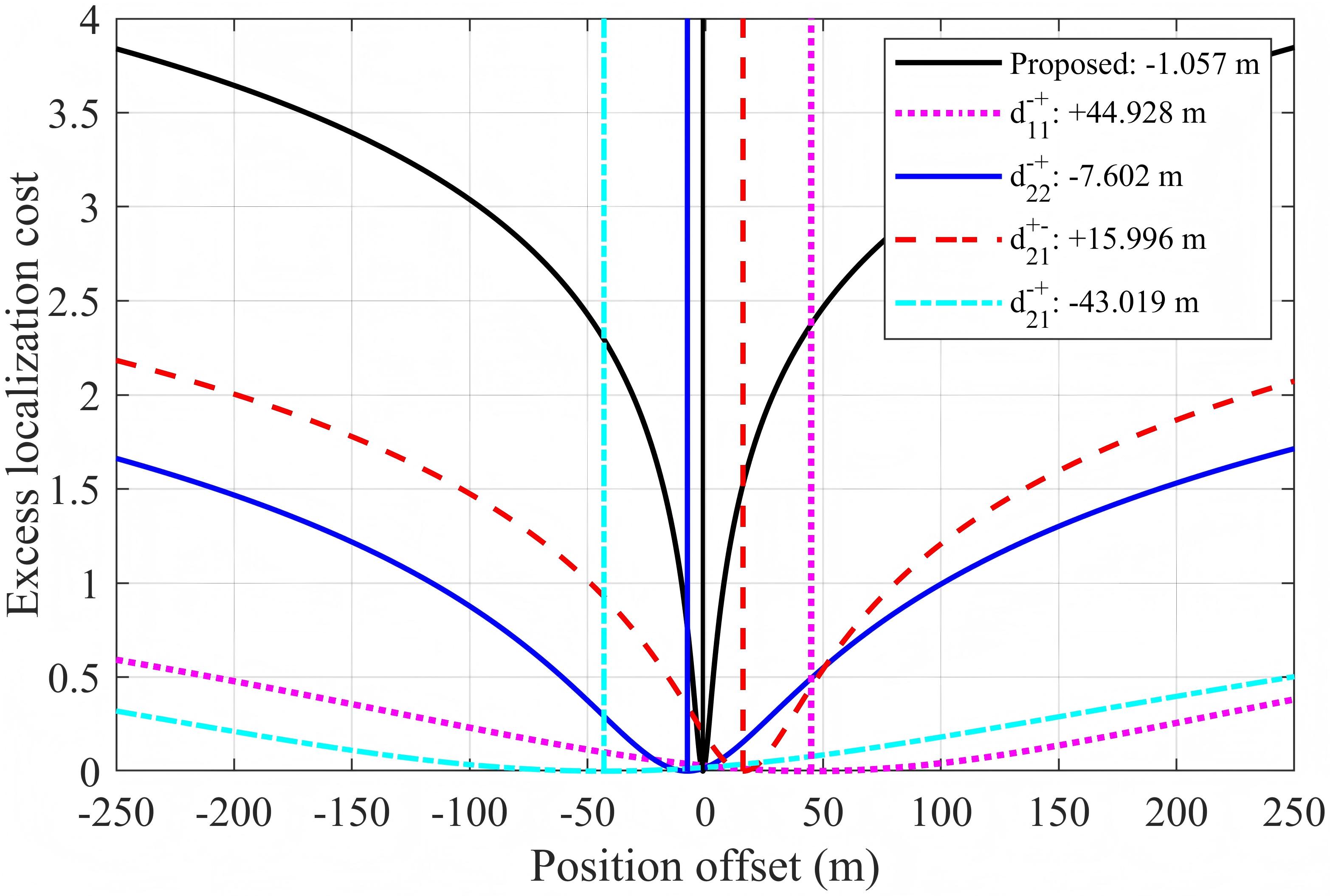}
    \caption{Localization-cost comparison between the proposed method and the four single-pair methods for the 500-mV record in the nominal 100-km group, where the ordinate is $\log_{10}(1+\Delta Q)$ and $\Delta Q$ is the increase above each curve's own minimum.}
    \label{Resfig:500mvLoc}
\end{figure}

Fig.~\ref{Resfig:500mvLoc} compares the proposed joint estimate with four single-pair estimates. The black curve corresponds to the joint cost, whereas the magenta, blue, red, and cyan curves correspond to $d_{11}^{-+}$, $d_{22}^{-+}$, $d_{21}^{+-}$, and $d_{21}^{-+}$, respectively. Each single-pair curve uses only its unwrapped delay and the same calibrated affine model:
\begin{equation}
Q_m^{\mathrm u}(z_{\mathrm{v}})
=\frac{(\widehat d_m^{\mathrm u}-a_m z_{\mathrm{v}}-b_m)^2}
{[\boldsymbol\Sigma_{\mathrm{uu}}]_{mm}},
\label{eq:experimental_single_pair_cost}
\end{equation}
where $m=3,\ldots,6$.
Its minimum gives $\widehat z_m=\Pi_{\mathcal Z}[(\widehat d_m^{\mathrm u}-b_m)/a_m]$, with $\mathcal Z=[0,L_{\mathrm{s}}]$. For the proposed method, the joint search in \eqref{eq:joint_integer_position_estimate} first determines $(\widehat z_{\mathrm{v}},\widehat{\boldsymbol n})$. The displayed black curve then holds $\widehat{\boldsymbol n}$ fixed and evaluates $Q(z_{\mathrm{v}},\widehat{\boldsymbol n})$. It consequently describes the selected integer-cycle branch, rather than the profile cost $Q_{\mathrm p}$ obtained by reoptimizing the integers at every position.

To display the different cost scales, the ordinate for the proposed method is
\begin{equation}
\mathcal E(z_{\mathrm{v}})
=\log_{10}\!\left(1+Q(z_{\mathrm{v}},\widehat{\boldsymbol n})
-Q(\widehat z_{\mathrm{v}},\widehat{\boldsymbol n})\right).
\label{eq:displayed_excess_cost}
\end{equation}
The same transformation is applied to each $Q_m^{\mathrm u}$ after subtracting its own minimum. This dimensionless logarithmic display preserves the minimizing positions and is not the relative score in \eqref{eq:relative_position_score}. The vertical markers indicate those positions. Zero cost on every curve results from separate minimum subtraction, not equal residual errors.

The horizontal coordinate is $z_{\mathrm{v}}-\overline z_{20}$, where $\overline z_{20}=101.018956$~km is the joint mean of 20 experimental records in this voltage--distance group. This reference $\overline z_{20}$ is used only for display and does not enter the localization search. The current joint estimate is $\widehat z_{\mathrm{v}}=101.017900$~km, corresponding to an offset of $-1.057$~m. The single-pair offsets are $+44.928$, $-7.602$, $+15.996$, and $-43.019$~m for $d_3$--$d_6$, respectively. These are deviations from a sample mean that includes the current record, not errors relative to an independently measured position.

The narrower black minimum indicates a larger local cost curvature under the adopted joint weighting. For this interior solution within the selected branch, \eqref{eq:joint_integer_costadd} and \eqref{eq:joint_integer_cost} give $\Delta Q=(\boldsymbol h^T\boldsymbol\Sigma^{-1}\boldsymbol h)(z_{\mathrm{v}}-\widehat z_{\mathrm{v}})^2$. The common-position model combines the six delay pairs, while the conditional term in \eqref{eq:conditional_cost_decomposition} incorporates the additional agreement of the restored wrapped observations without treating the two delay representations as independent. Thus, the joint minimum need not coincide with any single-pair minimum or their arithmetic mean. Under the adopted weighting, the narrower joint-cost minimum indicates a faster increase in cost as the candidate position moves away from the estimate within the selected integer-cycle branch.   

\begin{figure}[!t]
    \centering
    \includegraphics[width=0.48\textwidth]{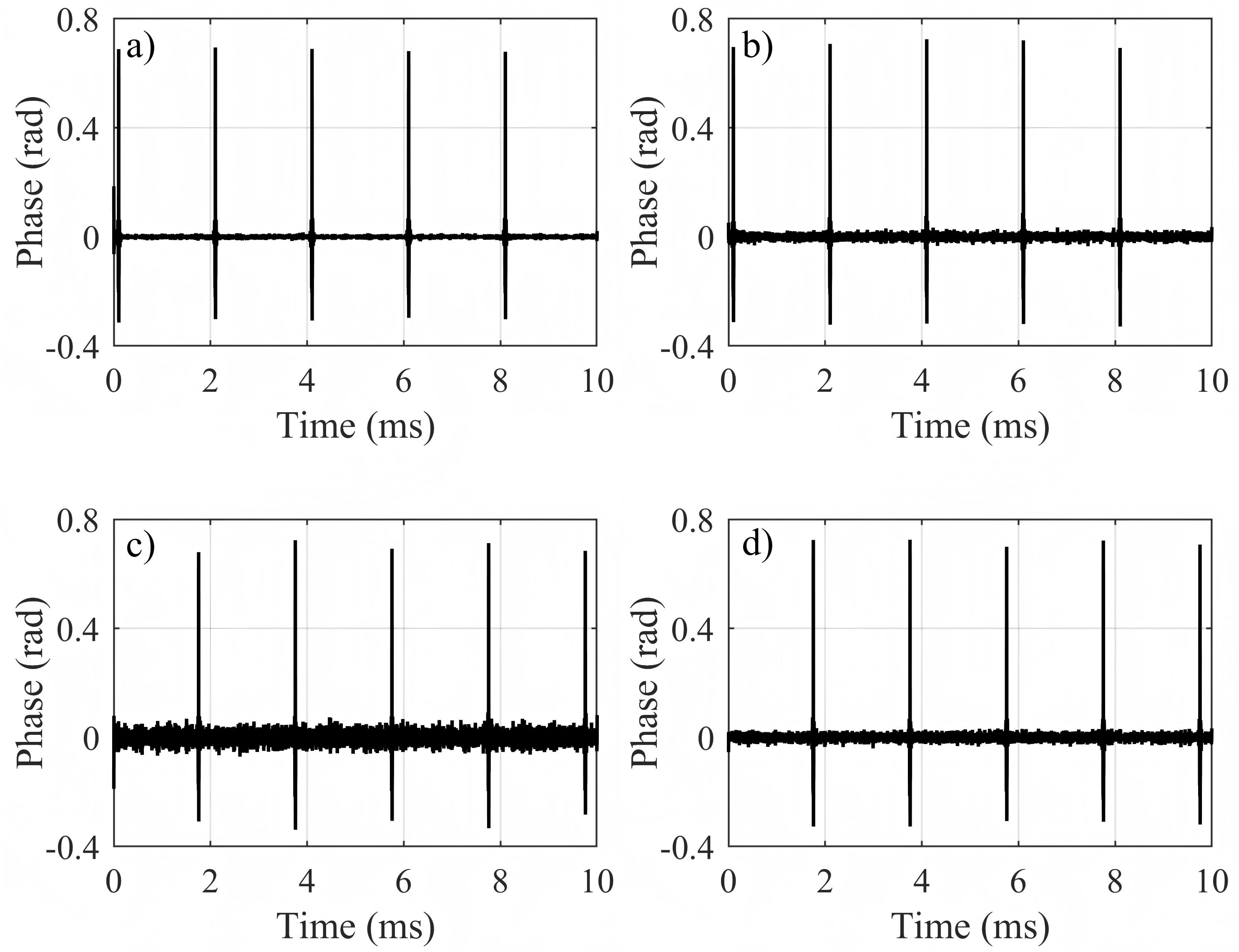}
    \caption{Recovered band-limited phase traces for a 1-V measurement in the nominal 100-km group: (a) $\widehat\varphi_1^{-}$, (b) $\widehat\varphi_2^{-}$, (c) $\widehat\varphi_1^{+}$, and (d) $\widehat\varphi_2^{+}$.}
    \label{Resfig:1vfour}
\end{figure}

Fig.~\ref{Resfig:1vfour} presents the phase traces at 1~V for the same nominal distance group, using the same channel order and processing as in Fig.~\ref{Resfig:500mvfour}. The positive phase peaks are approximately 0.7~rad, compared with approximately 0.35~rad at 500~mV, consistent with the increased PM drive. The localized features are more prominent relative to the background, although channel-dependent fluctuations remain visible, especially in the 1530-nm forward trace in panel~(c).

\begin{figure}[!t]
    \centering
    \includegraphics[width=0.46\textwidth]{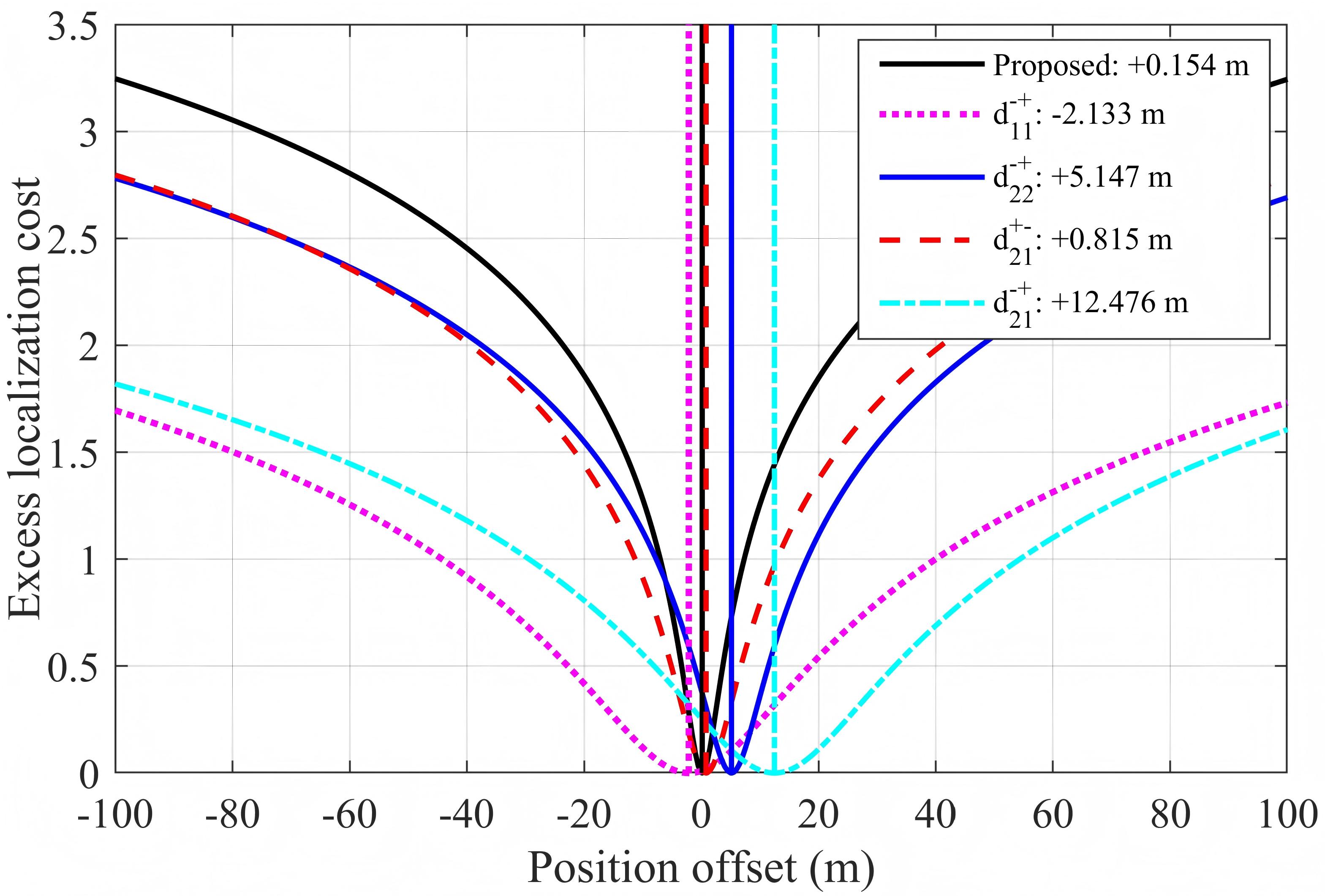}
    \caption{Localization-cost comparison between the proposed method and the four single-pair methods for the 1-V record in the nominal 100-km group, where the ordinate is $\log_{10}(1+\Delta Q)$ and $\Delta Q$ is the increase above each curve's own minimum.}
    \label{Resfig:1vLoc}
\end{figure}

Fig.~\ref{Resfig:1vLoc} shows the resulting localization costs, with zero referenced to the 20-record joint mean of the 1-V group. The proposed estimate has an offset of $+0.154$~m, whereas $d_3$--$d_6$ give $-2.133$, $+5.147$, $+0.815$, and $+12.476$~m, respectively. The maximum separation among these four single-pair estimates is 14.609~m, compared with 87.947~m for the 500-mV record in Fig.~\ref{Resfig:500mvLoc}. Thus, the single-pair results show closer mutual agreement in the illustrated 1-V record. Their minima nevertheless remain separated, and the proposed estimate is closer to its group mean than any of the four single-pair estimates. The smaller spread is consistent with the more prominent phase features, while the remaining pair-dependent differences are reconciled by the common-position fit through the joint delay weighting.

\begin{figure}[!t]
    \centering
    \includegraphics[width=0.42\textwidth]{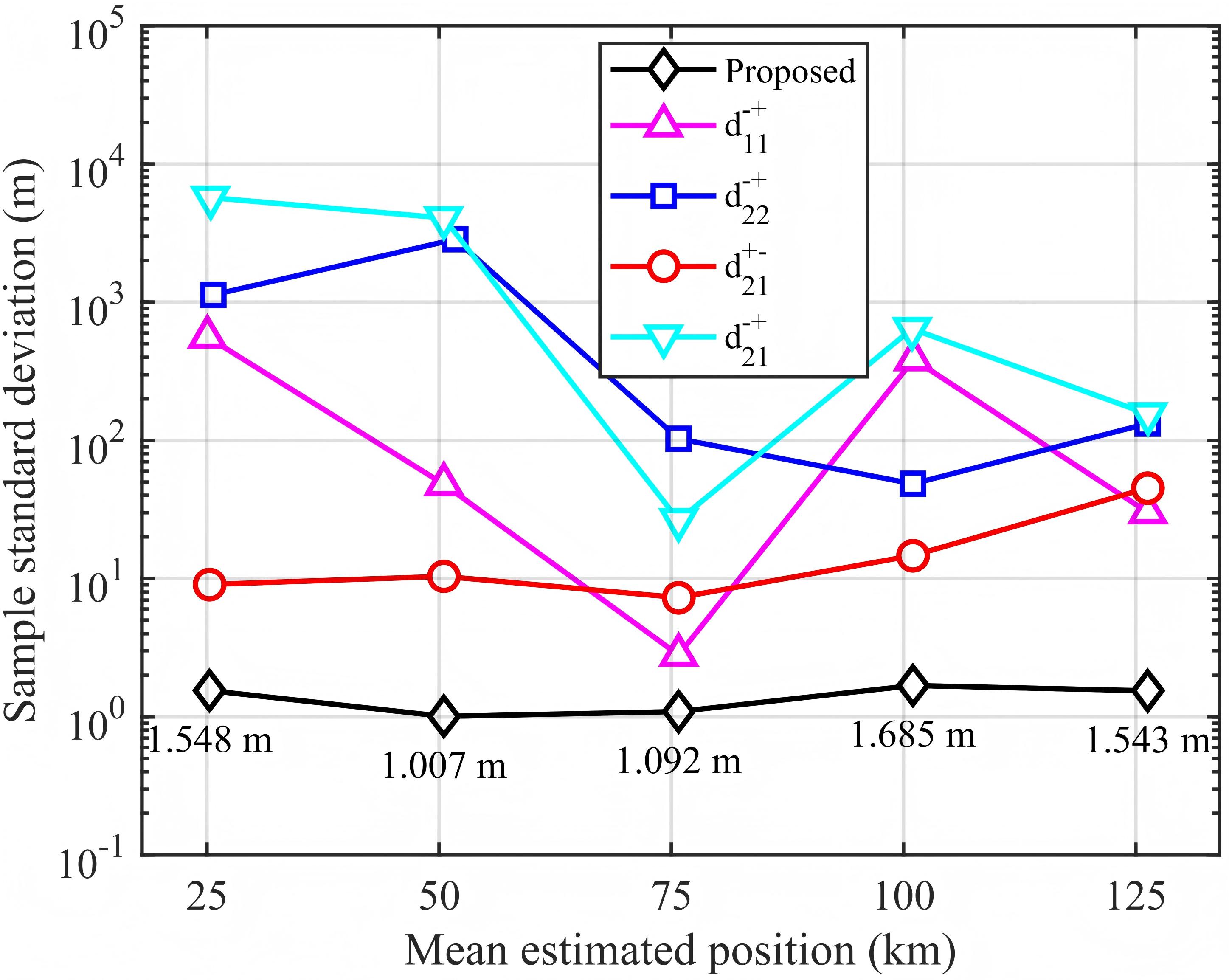}
    \caption{Localization repeatability of the proposed method and the four single-pair methods at a PM drive voltage of 500~mV, where the sample standard deviation is plotted against each method's mean estimated position.}
    \label{Resfig:500mvMean}
\end{figure}

\begin{figure}[!t]
    \centering
    \includegraphics[width=0.42\textwidth]{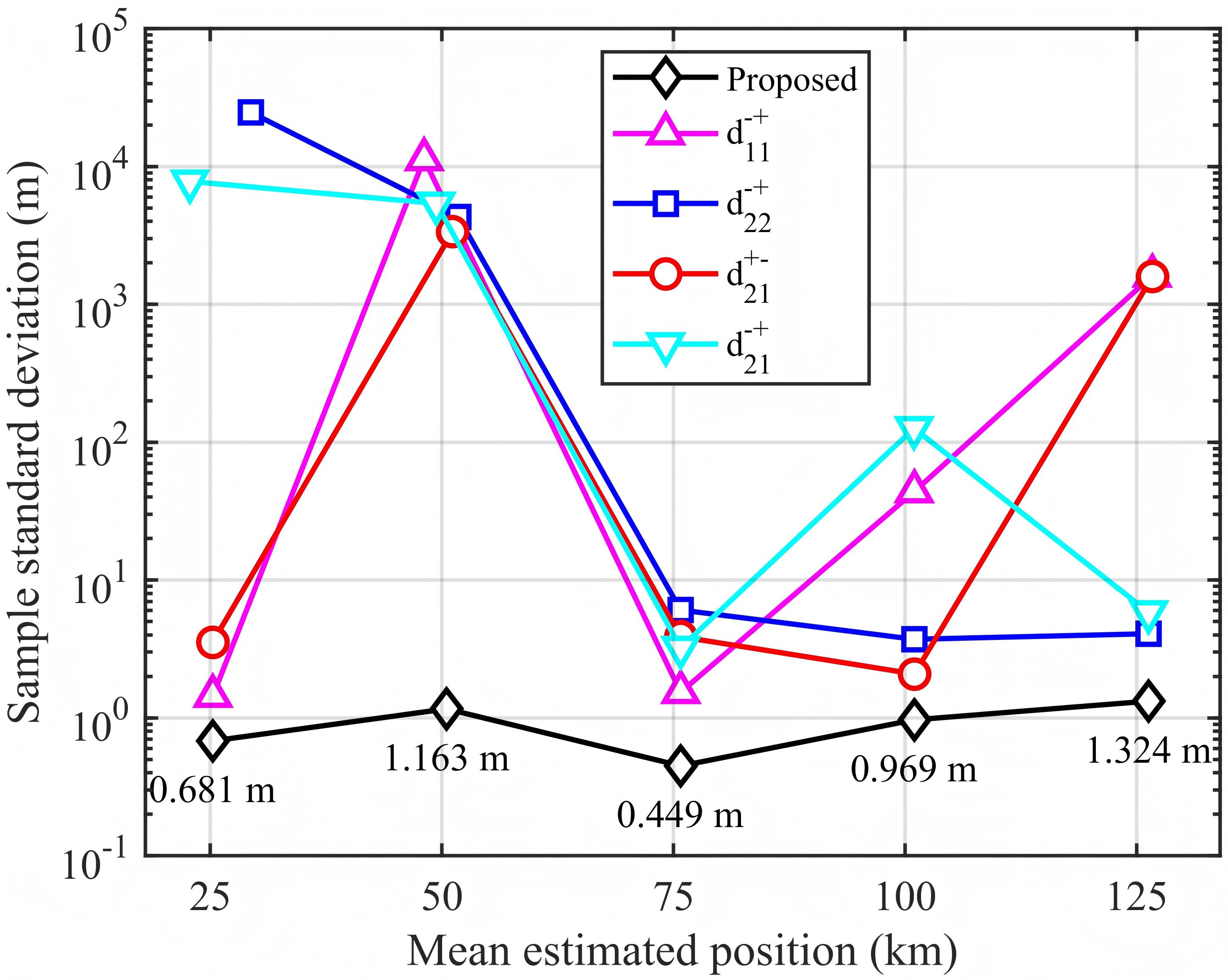}
    \caption{Localization repeatability of the proposed method and the four single-pair methods at a PM drive voltage of 1~V, where the sample standard deviation is plotted against each method's mean estimated position.}
    \label{Resfig:1vMean}
\end{figure}

Fig.~\ref{Resfig:500mvMean} and Fig.~\ref{Resfig:1vMean} extend the single-record comparisons to repeated measurements in the nominal 25-, 50-, 75-, 100-, and 125-km groups at 500~mV and 1~V, respectively. Each figure compares the proposed method with the four single-pair methods: the abscissa is each method's mean estimated position, and the ordinate is its sample standard deviation, shown in meters on a logarithmic scale. The standard deviations are listed in Tables~\ref{tab:repeatability_500mv} and~\ref{tab:repeatability_1v}.  All five methods use the same 20 experimental records. 

At 500~mV, the proposed method yields standard deviations of 1.007--1.685~m and has the smallest value in all five groups (Fig.~\ref{Resfig:500mvMean} and Table~\ref{tab:repeatability_500mv}). The smallest single-pair standard deviation in each group is 2.57--19.56 times that of the proposed method. Among the single-pair methods, $d_{21}^{+-}$ gives the lowest standard deviation at nominal distances of 25, 50, and 100~km, whereas $d_{11}^{-+}$ does so at 75 and 125~km. This variation cannot be explained by a large difference in position sensitivity: \eqref{eq:affine_coefficient1} gives $|a_m|\simeq s$ for all four counter-propagating pairs. For an interior single-pair solution, \eqref{eq:experimental_single_pair_cost} maps a delay error directly into a position error $\epsilon_{\mathrm u,m}/a_m$, without testing its compatibility with the other delays. The uneven single-pair repeatability therefore reflects differences in the measured delay errors, rather than substantially different delay-to-position gains.

At 1~V, the proposed standard deviations are 0.449--1.324~m, again the smallest in all five groups (Fig.~\ref{Resfig:1vMean} and Table~\ref{tab:repeatability_1v}). The proposed value decreases at 1~V in four groups; in the nominal 50-km group it increases from 1.007 to 1.163~m, while all four single-pair standard deviations exceed $3.3\times10^3$~m. The smallest single-pair value, 3367.519~m for $d_5=d_{21}^{+-}$, gives a ratio of approximately $2.90\times10^3$ to the proposed value. This large ratio is driven by a subset of records with large single-pair excursions. For example, in one retained record, $d_4$ and $d_5$ yield 67.719 and 65.227~km, respectively, whereas $d_3$, $d_6$, and the joint estimator yield 50.508, 50.499, and 50.507~km.

The single-pair baselines obtain position by directly inverting the unwrapped phase-slope delays in \eqref{eq:experimental_single_pair_cost}; their large excursions therefore occur before any joint integer search. In the above record, $d_4$ and $d_5$ have median spectral coherences of 0.223 and 0.269 and effective unwrapped-delay uncertainty scales above 100~$\mu\mathrm{s}$, whereas the corresponding uncertainty scales for $d_3$ and $d_6$ are below 0.13~$\mu\mathrm{s}$. These diagnostics indicate unstable delay estimation in the affected pairs despite the higher drive. The covariance-weighted common-position fit accounts for both the variation in delay reliability and the disagreement across pairs. A position displacement $\Delta z$ must change the full predicted delay vector by $\boldsymbol a\Delta z$ in \eqref{eq:joint_affine_model}, which the observed excursions confined to a subset of pairs do not follow.

For a fixed integer vector and weighting matrix, the interior solution in \eqref{eq:conditional_position_solution} shows that a perturbation of the stacked observations changes the fitted position only through its projection along $\boldsymbol h$ in the $\boldsymbol\Sigma^{-1}$ metric; orthogonal perturbations change the fitting residual rather than the position. The joint estimate therefore need not inherit the full excursions of an individual pair. Note that errors aligned with the physical direction can still shift the estimate. Thus, the constraint does not remove arbitrary delay errors.

Minimizing only the unwrapped term in \eqref{eq:conditional_cost_decomposition} over the same admissible position interval yields a sample standard deviation of 4.928~m on the same 20 records. Including the wrapped observations reduces this value to 1.163~m. This comparison shows that covariance-weighted fusion of the unwrapped delays already limits the influence of the single-pair excursions, with the wrapped observations providing a further improvement in this group.

\begin{table}[!t]
\centering
\caption{Sample Standard Deviation (m) at 500~mV}
\label{tab:repeatability_500mv}
\begin{threeparttable}
\footnotesize
\setlength{\tabcolsep}{3pt}
\renewcommand{\arraystretch}{1.25}
\begin{tabular*}{\columnwidth}{@{\extracolsep{\fill}} |l|c|c|c|c|c| @{}}
\hline
Group & 25 & 50 & 75 & 100 & 125 \\
\hline
Proposed & 1.548 & 1.007 & 1.092 & 1.685 & 1.543 \\
\hline
$d_{11}^{-+}$ only & 560.722 & 48.297 & 2.804 & 386.392 & 30.182 \\
\hline
$d_{22}^{-+}$ only & 1133.586 & 2859.666 & 102.713 & 48.674 & 132.319 \\
\hline
$d_{21}^{+-}$ only & 9.047 & 10.378 & 7.281 & 14.631 & 45.115 \\
\hline
$d_{21}^{-+}$ only & 5706.148 & 4055.204 & 26.460 & 640.847 & 154.996 \\
\hline
\end{tabular*}
\end{threeparttable}
\end{table}

\begin{table}[!t]
\centering
\caption{Sample Standard Deviation (m) at 1~V}
\label{tab:repeatability_1v}
\begin{threeparttable}
\footnotesize
\setlength{\tabcolsep}{3pt}
\renewcommand{\arraystretch}{1.25}
\begin{tabular*}{\columnwidth}{@{\extracolsep{\fill}}|l|c|c|c|c|c|@{}}
\hline
Group & 25 & 50 & 75 & 100 & 125 \\
\hline
Proposed & 0.681 & 1.163 & 0.449 & 0.969 & 1.324 \\
\hline
$d_{11}^{-+}$ only & 1.444 & 11324.400 & 1.542 & 44.002 & 1616.348 \\
\hline
$d_{22}^{-+}$ only & 24661.949 & 4312.838 & 6.079 & 3.718 & 4.074 \\
\hline
$d_{21}^{+-}$ only & 3.536 & 3367.519 & 3.892 & 2.078 & 1589.141 \\
\hline
$d_{21}^{-+}$ only & 7796.733 & 5396.414 & 3.231 & 126.545 & 5.848 \\
\hline
\end{tabular*}
\end{threeparttable}
\end{table}

The additional role of the wrapped observations follows from \eqref{eq:conditional_cost_decomposition}. The unwrapped term supplies cross-branch support from the measured spectral phase slope, whereas $\boldsymbol e_{\mathrm r}-\mathbf K\boldsymbol e_{\mathrm u}$ tests the restored wrapped delays after accounting for their modeled correlation with the unwrapped errors. A position displacement changes this conditional residual by $-(\boldsymbol a-\mathbf K\boldsymbol a)\Delta z$; when this sensitivity is nonzero, the wrapped observations add a position-dependent penalty beyond the unwrapped fit. Joint estimation therefore provides a mechanism for combining resistance to cross-pair inconsistency with supplementary phase-based position sensitivity. The full covariance accounts for shared-channel and cross-representation dependence, rather than treating the delay estimates as independent measurements. 

For near-common aliases whose dispersion-scale differences in \eqref{eq:near_common_position_alias} are small relative to the delay errors, reliable unwrapped observations provide complementary branch discrimination that the wrapped constraints alone may not reliably supply. Thus, the physical model specifies the admissible delay combinations, while the complementary spectral observations and their joint weighting determine the selected position within that model.

\section{Conclusion}\label{Conclu}
We developed a novel spectral delay fusion scheme based on a one-dimensional affine model for a distributed optical-fiber sensing system using a dual-wavelength bidirectional MZI. The method jointly estimates the position of a single dominant disturbance and integer ambiguities from correlated unwrapped and wrapped delay observations.
The six pairwise delays provide position responses governed by chromatic dispersion and propagation direction. With propagation parameters and timing offsets fixed, their predictions form a one-dimensional affine line segment. The common-position fit tests delay consistency: a deviation confined to a subset of pairs cannot generally be explained by one position displacement. Errors aligned with the model direction can still shift the estimate.

The cost decomposition separates the unwrapped-delay fit from the conditional contribution of the restored wrapped delays. Reliable spectral phase slopes support discrimination among periodic candidates, while the wrapped delays can add phase sensitivity after accounting for their correlation with the unwrapped estimates. The effective joint covariance accounts for dependence introduced by shared channels and reference-frequency alignment.
For periodic disturbances, exact noiseless identifiability and a narrow cost minimum for a fixed integer ambiguity vector do not guarantee correct position selection. Candidate positions can remain difficult to distinguish when their dispersion-induced delay differences are small relative to measurement errors. Local precision and correct integer-ambiguity selection therefore remain distinct performance requirements.

In the 131.335-km experiment, the reported 20-record groups at five nominal positions from 25 to 125~km yielded sample standard deviations of 1.007--1.685~m at 500~mV and 0.449--1.324~m at 1~V. The ratio of the smallest single-pair sample standard deviation to that of the proposed method ranged from 2.57 to 19.56 at 500~mV and from 2.12 to $2.90\times10^3$ at 1~V. The upper ratio at 1~V arose in the nominal 50-km group, where large excursions in individual phase-slope delay estimates increased the smallest single-pair standard deviation to 3367.519~m, while the joint estimate retained a standard deviation of 1.163~m. These statistics quantify localization repeatability; the coarse OTDR references do not verify absolute position errors at the same scale.
Further work should extend the model to simultaneous disturbances and assess calibration stability under changing link conditions. Independent position references with sufficient accuracy are also needed to evaluate absolute localization accuracy.


\begin{thebibliography}{99}
\urlstyle{same}

\bibitem{Bao2017}
X.~Bao, D.-P.~Zhou, C.~Baker, and L.~Chen,
``Recent development in the distributed fiber optic acoustic and ultrasonic detection,''
\emph{J. Lightw. Technol.}, vol.~35, no.~16, pp.~3256--3267, 2017,
doi:~\nolinkurl{10.1109/JLT.2016.2612060}.

\bibitem{Lindsey2019}
N. J. Lindsey, T. C. Dawe, and J. B. Ajo-Franklin, ``Illuminating seafloor faults and ocean dynamics with dark fiber distributed acoustic sensing,'' \emph{Science}, vol.~366, no.~6469, pp.~1103--1107, 2019, doi:~\nolinkurl{10.1126/science.aay5881}.

\bibitem{Sladen2019}
A. Sladen \emph{et al.}, ``Distributed sensing of earthquakes and ocean-solid Earth interactions on seafloor telecom cables,'' \emph{Nat. Commun.}, vol.~10, no.~1, Art. no.~5777, 2019, doi:~\nolinkurl{10.1038/s41467-019-13793-z}.

\bibitem{Huang2026CPOFDM}
H.~Huang \emph{et al.},
``Cyclic-prefix OFDM probing for spatial-ISI-free distributed acoustic sensing via frequency-domain channel reconstruction,''
arXiv preprint arXiv:2606.19724, 2026,
doi:~\nolinkurl{10.48550/arXiv.2606.19724}.

\bibitem{Huang2026DMT}
H.~Huang, Z.~Chen, Z.~Xue, D.~Zou, and Y.~Cai,
``A continuous payload-bearing discrete multitone modulation framework for fiber-optic integrated sensing and communication,''
arXiv preprint arXiv:2608.29020, 2026,
doi:~\nolinkurl{10.48550/arXiv.2608.29020}.

\bibitem{Marra2018}
G. Marra \emph{et al.}, ``Ultrastable laser interferometry for earthquake detection with terrestrial and submarine cables,'' \emph{Science}, vol.~361, no.~6401, pp.~486--490, 2018, doi:~\nolinkurl{10.1126/science.aat4458}.

\bibitem{Yan2021}
Y. Yan, F. N. Khan, B. Zhou, A. P. T. Lau, C. Lu, and C. Guo, ``Forward transmission based ultra-long distributed vibration sensing with wide frequency response,'' \emph{J. Lightw. Technol.}, vol.~39, no.~7, pp.~2241--2249, 2021, doi:~\nolinkurl{10.1109/JLT.2020.3044676}.

\bibitem{Mazur2022}
M. Mazur \emph{et al.}, ``Transoceanic phase and polarization fiber sensing using real-time coherent transceiver,'' in \emph{Proc. Opt. Fiber Commun. Conf. (OFC)}, 2022, Paper~M2F.2, doi:~\nolinkurl{10.1364/OFC.2022.M2F.2}.

\bibitem{Marra2022}
G. Marra \emph{et al.}, ``Optical interferometry-based array of seafloor environmental sensors using a transoceanic submarine cable,'' \emph{Science}, vol.~376, no.~6595, pp.~874--879, 2022, doi:~\nolinkurl{10.1126/science.abo1939}.

\bibitem{Hong2011}
X. Hong, J. Wu, C. Zuo, F. Liu, H. Guo, and K. Xu, ``Dual Michelson interferometers for distributed vibration detection,'' \emph{Appl. Opt.}, vol.~50, no.~22, pp.~4333--4338, 2011, doi:~\nolinkurl{10.1364/AO.50.004333}.

\bibitem{Hoffman2004}
P. R. Hoffman and M. G. Kuzyk, ``Position determination of an acoustic burst along a Sagnac interferometer,'' \emph{J. Lightw. Technol.}, vol.~22, no.~2, pp.~494--498, 2004, doi:~\nolinkurl{10.1109/JLT.2004.824455}.

\bibitem{Teng2019}
F. Teng, D. Yi, X. Hong, and X. Li, ``Distributed fiber optics disturbance sensor using a dual-Sagnac interferometer,'' \emph{Opt. Lett.}, vol.~44, no.~20, pp.~5101--5103, 2019, doi:~\nolinkurl{10.1364/OL.44.005101}.

\bibitem{Hu2021}
Y. Hu \emph{et al.}, ``An asymmetrical dual Sagnac distributed fiber sensor for high precision localization based on time delay estimation,'' \emph{J. Lightw. Technol.}, vol.~39, no.~21, pp.~6928--6933, 2021, doi:~\nolinkurl{10.1109/JLT.2021.3105299}.

\bibitem{Teng2021}
F. Teng, D. Yi, X. Hong, and X. Li, ``Optimized localization algorithm of dual-Sagnac structure-based fiber optic distributed vibration sensing system,'' \emph{Opt. Express}, vol.~29, no.~9, pp.~13696--13705, 2021, doi:~\nolinkurl{10.1364/OE.421569}.

\bibitem{Spammer1997}
S. J. Spammer, P. L. Swart, and A. A. Chtcherbakov, ``Merged Sagnac--Michelson interferometer for distributed disturbance detection,'' \emph{J. Lightw. Technol.}, vol.~15, no.~6, pp.~972--976, 1997, doi:~\nolinkurl{10.1109/50.588669}.

\bibitem{Song2020}
Q. Song \emph{et al.}, ``Improved localization algorithm for distributed fiber-optic sensor based on merged Michelson--Sagnac interferometer,'' \emph{Opt. Express}, vol.~28, no.~5, pp.~7207--7220, 2020, doi:~\nolinkurl{10.1364/OE.384728}.

\bibitem{Liang2009}
S. Liang \emph{et al.}, ``Fiber-optic intrinsic distributed acoustic emission sensor for large structure health monitoring,'' \emph{Opt. Lett.}, vol.~34, no.~12, pp.~1858--1860, 2009, doi:~\nolinkurl{10.1364/OL.34.001858}.

\bibitem{DiLuch2021}
I. Di Luch, P. Boffi, M. Ferrario, G. Rizzelli, R. Gaudino, and M. Martinelli, ``Vibration sensing for deployed metropolitan fiber infrastructure,'' \emph{J. Lightw. Technol.}, vol.~39, no.~4, pp.~1204--1211, 2021, doi:~\nolinkurl{10.1109/JLT.2021.3051732}.

\bibitem{Chen2014WalkOff}
Q. Chen \emph{et al.}, ``A distributed fiber vibration sensor utilizing dispersion induced walk-off effect in a unidirectional Mach--Zehnder interferometer,'' \emph{Opt. Express}, vol.~22, no.~3, pp.~2167--2173, 2014, doi:~\nolinkurl{10.1364/OE.22.002167}.

\bibitem{Ma2016ADMZI}
C. Ma \emph{et al.}, ``Long-range distributed fiber vibration sensor using an asymmetric dual Mach--Zehnder interferometers,'' \emph{J. Lightw. Technol.}, vol.~34, no.~9, pp.~2235--2239, 2016, doi:~\nolinkurl{10.1109/JLT.2016.2532877}.

\bibitem{Chen2023OL}
G. Y. Chen \emph{et al.}, ``Long-range distributed vibration sensing using phase-sensitive forward optical transmission,'' \emph{Opt. Lett.}, vol.~48, no.~18, pp.~4825--4828, 2023, doi:~\nolinkurl{10.1364/OL.500587}.

\bibitem{Rao2024JLT}
X. Rao, Y. Wang, M. Chen, K. Liu, G. Y. Chen, and Y. Wang, ``150 km single-span distributed vibration sensor based on compensated self-interference forward transmission,'' \emph{J. Lightw. Technol.}, vol.~42, no.~16, pp.~5736--5742, 2024, doi:~\nolinkurl{10.1109/JLT.2024.3397782}.

\bibitem{Quazi1981}
A.~H.~Quazi,
``An overview on the time delay estimate in active and passive systems for target localization,''
\emph{IEEE Trans. Acoust., Speech, Signal Process.}, vol.~29, no.~3, pp.~527--533, 1981,
doi:~\nolinkurl{10.1109/TASSP.1981.1163618}.

\bibitem{Ma2024}
B. Ma, R. Jin, C. Li, Y. Wu, C. Wang, and B. Jia, ``Improved vibration localization algorithm for multiple intrusions based on phase spectrum estimation in distributed Mach--Zender/Sagnac optical fiber sensing system,'' \emph{IEEE Sensors J.}, vol.~24, no.~8, pp.~12426--12432, 2024, doi:~\nolinkurl{10.1109/JSEN.2024.3372648}.

\bibitem{Rao2024OE}
X. Rao \emph{et al.}, ``Multi-point vibration positioning method for long-distance forward transmission distributed vibration sensing,'' \emph{Opt. Express}, vol.~32, no.~17, pp.~30775--30786, 2024, doi:~\nolinkurl{10.1364/OE.530885}.

\bibitem{Jin2025}
R. Jin \emph{et al.}, ``Positioning error limits and noise analysis in hybrid MZ--Sagnac interferometric distributed optical fiber sensing system,'' \emph{J. Lightw. Technol.}, vol.~43, no.~5, pp.~2438--2448, 2025, doi:~\nolinkurl{10.1109/JLT.2024.3486815}.

\bibitem{Fang2026}
J. Fang, Y. Li, W. Kohno, and T. Wang, ``High-fidelity forward-transmission vibration sensing using dual-sensor beamforming,'' \emph{J. Lightw. Technol.}, vol.~44, no.~16, pp.~7265--7273, 2026, doi:~\nolinkurl{10.1109/JLT.2026.3701949}.

\bibitem{Knapp1976}
C.~H.~Knapp and G.~C.~Carter,
``The generalized correlation method for estimation of time delay,''
\emph{IEEE Trans. Acoust., Speech, Signal Process.}, vol.~24, no.~4, pp.~320--327, 1976,
doi:~\nolinkurl{10.1109/TASSP.1976.1162830}.

\bibitem{Benesty2004}
J.~Benesty, J.~Chen, and Y.~Huang,
``Time-delay estimation via linear interpolation and cross correlation,''
\emph{IEEE Trans. Speech Audio Process.}, vol.~12, no.~5, pp.~509--519, 2004,
doi:~\nolinkurl{10.1109/TSA.2004.833008}.

\bibitem{Ianniello1982}
J.~P.~Ianniello,
``Time delay estimation via cross-correlation in the presence of large estimation errors,''
\emph{IEEE Trans. Acoust., Speech, Signal Process.}, vol.~30, no.~6, pp.~998--1003, 1982,
doi:~\nolinkurl{10.1109/TASSP.1982.1163992}.

\bibitem{Akhlaq2016Wavelengths}
A.~Akhlaq, R.~G.~McKilliam, R.~Subramanian, and A.~Pollok,
``Selecting wavelengths for least squares range estimation,''
\emph{IEEE Trans. Signal Process.}, vol.~64, no.~20, pp.~5205--5216, 2016,
doi:~\nolinkurl{10.1109/TSP.2016.2595490}.

\bibitem{Hassibi1998Integer}
A.~Hassibi and S.~Boyd,
``Integer parameter estimation in linear models with applications to GPS,''
\emph{IEEE Trans. Signal Process.}, vol.~46, no.~11, pp.~2938--2952, 1998,
doi:~\nolinkurl{10.1109/78.726808}.

\bibitem{Tucker2023MixedInteger}
D.~Tucker, S.~Zhao, and L.~C.~Potter,
``Maximum likelihood estimation in mixed integer linear models,''
\emph{IEEE Signal Process. Lett.}, vol.~30, pp.~1557--1561, 2023,
doi:~\nolinkurl{10.1109/LSP.2023.3324833}.

\bibitem{Kikuchi2016}
K.~Kikuchi,
``Fundamentals of coherent optical fiber communications,''
\emph{J. Lightw. Technol.}, vol.~34, no.~1, pp.~157--179, 2016,
doi:~\nolinkurl{10.1109/JLT.2015.2463719}.

\bibitem{Itoh1982}
K.~Itoh,
``Analysis of the phase unwrapping algorithm,''
\emph{Appl. Opt.}, vol.~21, no.~14, p.~2470, 1982,
doi:~\nolinkurl{10.1364/AO.21.002470}.

\bibitem{Welch1967}
P.~D.~Welch,
``The use of fast Fourier transform for the estimation of power spectra: A method based on time averaging over short, modified periodograms,''
\emph{IEEE Trans. Audio Electroacoust.}, vol.~15, no.~2, pp.~70--73, 1967,
doi:~\nolinkurl{10.1109/TAU.1967.1161901}.

\bibitem{Carter1987}
G.~C.~Carter,
``Coherence and time delay estimation,''
\emph{Proc. IEEE}, vol.~75, no.~2, pp.~236--255, 1987,
doi:~\nolinkurl{10.1109/PROC.1987.13723}.

\bibitem{Huber1964}
P.~J.~Huber,
``Robust estimation of a location parameter,''
\emph{Ann. Math. Statist.}, vol.~35, no.~1, pp.~73--101, 1964,
doi:~\nolinkurl{10.1214/aoms/1177703732}.

\bibitem{Teunissen1995}
P.~J.~G.~Teunissen,
``The least-squares ambiguity decorrelation adjustment: A method for fast GPS integer ambiguity estimation,''
\emph{J. Geodesy}, vol.~70, nos.~1--2, pp.~65--82, 1995,
doi:~\nolinkurl{10.1007/BF00863419}.

\end{thebibliography}
\end{document}